\RequirePackage{docswitch}				
\setjournal{\flag}						

\documentclass[12pt]{lsstdescnote}

\usepackage{soul} 
\usepackage{amsmath}
\usepackage{amssymb}
\usepackage{xspace}
\usepackage{xifthen}
\usepackage[dvipsnames,svgnames]{xcolor} 

\mathchardef\mhyphen="2D

\newlength{\dhatheight}

\usepackage[outdir=./]{epstopdf}
\usepackage{graphicx}
\usepackage{color}
\usepackage{amsmath}
\usepackage{longtable}

\usepackage[10pt]{extsizes}

\usepackage{hyperref}
\hypersetup{
    colorlinks=true, 
    linktoc=all,     
}

\usepackage{array}

\graphicspath{{./}{./figures/}}
\usepackage[top=1.2in, bottom=0.1in, left=1.5in, right=1in,includehead, includefoot]{geometry}

\newcommand{\nn}{\nonumber}

\newcommand\ees{\end{eqnarray}}
\newcommand\bees{\begin{eqnarray}}

\newcommand{\be}{\begin{equation}}
\newcommand{\ee}{\end{equation}}
\newcommand{\bea}{\begin{eqnarray}}
\newcommand{\eea}{\end{eqnarray}}

\newcommand{\DD}{{\alpha}}

\newcommand{\Smn}{S_{\mu\nu}}

\renewcommand\({\left(}
\renewcommand\){\right)}
\renewcommand\[{\left[}
\renewcommand\]{\right]}

\newcommand\n{{\mbox {\boldmath $\nabla$}}}

\def\lsim{\raise 0.4ex\hbox{$<$}\kern -0.8em\lower 0.62
ex\hbox{$\sim$}}

\newcommand{\iBox}{\Box^{-1}}
\newcommand{\Tmn}{T_{\mu\nu}}
\newcommand{\gmn}{g_{\mu\nu}}

\newcommand{\Gmn}{G_{\mu\nu}}
\newcommand\eqst[2]{eqs.~(\ref{#1})--(\ref{#2})}

\newcommand{\ode}{\Omega_{\rm DE}}

\newcommand{\rde}{\rho_{\rm DE}}

\newcommand\eq[1]{eq.~(\ref{#1})}

\def\g{\gamma}

\def\d{\delta}

\begin{document}


\pagestyle{empty}

\vspace*{0.2\textheight}

\begin{center}
{\Huge\bfseries Modified Gravity and Dark Energy models Beyond $w(z)$CDM Testable by LSST}

\vspace*{0.2\textheight}

{\Large\bfseries Version~1.1}

Date: September 6, 2019 

\vspace*{0.1\textheight}

\begin{figure}[!h]
\centering\includegraphics[width=5cm,angle=0]{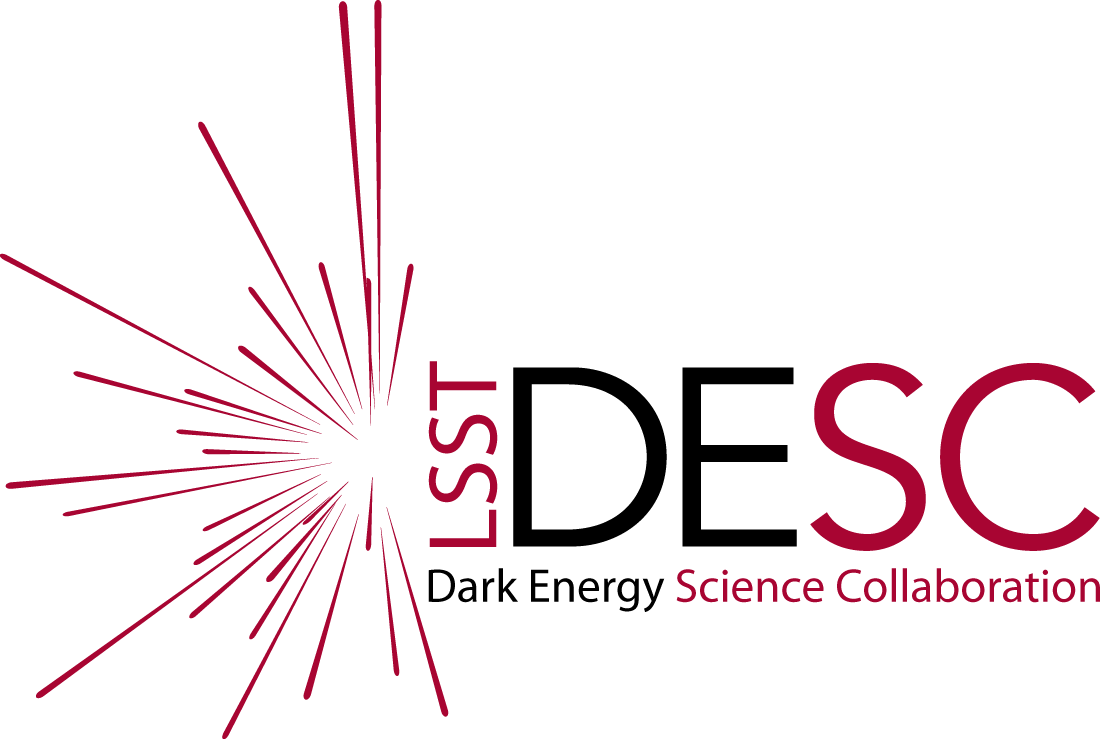}
\end{figure}

\end{center}

\clearpage

\newpage
\section*{Change Record}

\begin{table}[!thp]
\begin{center}

\renewcommand{\arraystretch}{1.2}
\begin{tabular}{|p{1.5cm}|p{2cm}|p{7.cm}|p{4cm}|}
\hline
{\bf Version}  &{\bf Date}  &{\bf Description}  & {\bf Owner name}
\\ \hline
v0.9   & 04/22/2019 & version for collaboration feedback.  & Mustapha Ishak
\\ \hline
v1.0   & 05/23/2019 & version after collaboration feedback.  & Mustapha Ishak
 \\ \hline
v1.1   & 09/06/2019 & some acknowledgments and references added.  & Mustapha Ishak
 \\ \hline
  &
  &
  &
\\ \hline
  &
  &
  &
\\ \hline
\end{tabular}
\renewcommand{\arraystretch}{1.0}
\end{center}
\end{table}

\newpage

\noindent Contributions to this note effort are listed in the table below in alphabetical order.
\vspace{-3mm}
\begin{longtable}{|p{4cm}|p{11cm}|}
\hline
{\bf Tessa Baker$^{1,2}$}   &{\bf Corresponding author.} Wrote the Horndeski and bigravity sections (together with P.Bull and P.Ferreira) sections 2.1 and 2.2. Edited and reviewed much of the document. Collated the mathematics and derived quasistatic limits for Horndeski equations. Determined rankings of priority models according to the scheme of section 4. t.baker@qmul.ac.uk\\ 
Jonathan Blazek$^{3,4}$ & Coordination and planning for document and model selection; wrote text on systematic effects; general editing.\\ 
{\bf Philip Bull$^{2,5}$}&{\bf Corresponding author.} Coordinated initial work on assessing models; helped coordinate rest of project; initial writing and organization of document; general editing. p.bull@qmul.ac.uk\\       
Pedro G. Ferreira$^1$ & Wrote text\\ 
{\bf Mustapha Ishak$^6$}&{\bf Lead and corresponding author}. Co-initiated and co-led the effort and writing of this note. Coordinated and contributed to work on assessing and ranking models. Contributed to a thorough compilation of MG literature and the writing of precursory text and tables on DESC confluence page that were used for this note. Wrote most of section 2.3 on Non-local Gravity and section 4 on Systematics; Contributed to writing sections 1.1 and 1.2; Edited and reviewed all the document. mishak@utdallas.edu\\                   
C. Danielle Leonard$^7$& Wrote text. Reviewed text. Participated in discussions\\
Weikang Lin$^6$        & Wrote text.\\ 
Eric Linder$^8$         & Initial ranking formalism; detailed discussions; thorough editing\\ 				 
Kris Pardo$^{9}$			 & Wrote sentences and edited Sections 2.1 and Appendix A\\
{\bf Eske M. Pedersen$^{6}$}& {\bf Corresponding author.} Wrote text for Sections 2.4 and 2.5 addressed related comments and edited both of these sections as well. emp160330@utdallas.edu\\    
Georgios Valogiannis$^{10}$ & Wrote text\\ 
\hline
\end{longtable}

{\small
\vspace{-3mm}
\noindent $^1$ Department of Physics, University of Oxford, Keble Road, Oxford OX1 3RH, UK\\
$^2$ School of Physics \& Astronomy, Queen Mary University of London, London E1 4NS, UK\\
$^3$ Center for Cosmology and Astroparticle Physics, Ohio State, Columbus, OH 43210, USA\\
$^4$ Laboratory of Astrophysics, E´cole Polytechnique Fe´de´ rale de Lausanne (EPFL), Observatoire de Sauverny, 1290 Versoix, Switzerland\\
$^5$ Radio Astronomy Laboratory, University of California Berkeley, Berkeley, CA 94720, USA\\
$^6$ Department of Physics, The University of Texas at Dallas, Richardson, TX 75083, USA\\
$^7$ McWilliams Center for Cosmology, Department of Physics, Carnegie Mellon University\\
$^8$ Berkeley Lab and University of California, Berkeley CA 94720 USA\\
$^9$ Department of Astrophysical Sciences, Princeton University, Princeton, NJ 08544, USA\\
$^{10}$ Department of Astronomy, Cornell University, Ithaca, NY 14853, USA\\}

\newpage

\clearpage

\title{Modified Gravity and Dark Energy models Beyond $w(z)$CDM Testable by LSST\vspace{-5mm}}

\maketitlepre				

\begin{abstract}
\vspace{-8mm}
One of the main science goals of the Large Synoptic Survey Telescope (LSST) is to uncover the nature of cosmic acceleration. In the base analysis, possible deviations from the Lambda-Cold-Dark-Matter ($\Lambda$CDM) background evolution will be probed by fitting a $w(z)$CDM model, which allows for a redshift-dependent dark energy equation of state with $w(z)$, within general relativity (GR). A rich array of other phenomena can arise due to deviations from the standard $\Lambda$CDM+GR model though, including modifications to the growth rate of structure and lensing, and novel screening effects on non-linear scales. Concrete physical models are needed to provide consistent predictions for these (potentially small) effects, to give us the best chance of detecting them and separating them from astrophysical systematics. A complex plethora of possible models has been constructed over the past few decades, with none emerging as a particular favorite. This document prioritizes a subset of these models along with rationales for further study and inclusion into the LSST Dark Energy Science Collaboration (DESC) data analysis pipelines, based on their observational viability, theoretical plausibility, and level of theoretical development. We provide references and theoretical expressions to aid the integration of these models into DESC software and simulations, and give justifications for why other models were not prioritized. While DESC efforts are free to pursue other models, we provide here guidelines on which theories appear to have higher priority for collaboration efforts due to their perceived promise and greater instructional value. 
\end{abstract}


\maketitlepost			


\tableofcontents{}
\newpage

\pagestyle{plain}

\section{Introduction}
\label{sec:intro} 

The purpose of this document is to identify a small selection of well-motivated modified gravity (MG) models to be used as the basis for developing a DE/MG analysis for the Dark Energy Science Collaboration (DESC) of the Large Synoptic Survey Telescope (LSST). This document provides a deliverable to the DESC science road map (SRM\footnote{\url{http://lsstdesc.org/sites/default/files/DESC\_SRM\_V1\_4.pdf}}) objective TJP2.1 of the Theory and Joint Probes (TJP) working group. The construction of models beyond GR/$w(z)$CDM is a major focus of the theoretical cosmology community, motivated by issues such as the cosmological constant problem, the observation of cosmic acceleration, predictions of new fields/particles from high-energy theory, and the fact that GR has not yet been tested precisely on cosmological scales. As a result, a large set of models exists in the literature, but not all are well-motivated or developed enough to be worth the extensive effort of testing them with LSST data. Equally importantly, the development of DE/MG analysis tools beyond tests of $w(z)$CDM and phenomenological treatments will also broaden the fundamental cosmology science return from LSST. Recent literature shows the tremendous interest from the scientific community in probing deviations from General Relativity at cosmological scales, see e.g., \cite{2012-Clifton-MG,2015-rev-Joyce-et-al,Cosmo.test.of.GR,KOYAMA2016TestGR,2016-Joyce-Lombriser-Schmidt-DEvsMG,Ishak2018}.

Specific examples of modified gravity models are needed to inform the design of an effective MG/DE analysis. When the assumptions behind GR are broken, or new types of field are added, new gravitational effects such as screening mechanisms can be introduced. The existence or behavior of such effects have been discovered or better understood by studying specific theories. While reliable phenomenological descriptions can now be made for some subset of the possibilities (e.g., the effective equation of state of dark energy, $w(z)$), these typically work within a certain framework, and can miss degrees of freedom outside that framework.  

Novel effects such as screening are also intrinsically non-linear in nature. Progress has been made on pushing parametrizations into the mildly non-linear regime \citep{HS2007PPF,Koyama2009N,Lombriser2014c,Lombriser2014b,Taruya2016,Nishimichi2017,Namikawa2018} and even into the deeply nonlinear regime \citep{Lombriser2016NL}; however, the study of N-body simulations of specific modified gravity models is needed in order to validate and improve such parametrizations \citep{Koyama2009N,Zhao2011Z}. Again, N-body simulations can cover models that are not within the framework of such parametrizations. Therefore studying some specific modified gravity models complements well phenomenological approaches.

Although a large number of modified gravity theories have been proposed in the literature (for example, see the reviews by \citep{2012-Clifton-MG,WillReview2014,2015-rev-Joyce-et-al,BertiElAlReview2015,KOYAMA2016TestGR,Ishak2018}, this study focuses on models that can have effects on cosmological scales and, in particular, can produce the late-time observed cosmic acceleration without a cosmological constant. It is also worth clarifying that even if modified gravity models do not solve the cosmological constant (vacuum energy) problems, they may still provide some hints on how to re-approach these problems. 
Indeed, before the discovery of cosmic acceleration, it was well-known that theoretical expectations for the cosmological constant exceed observational constraints by 120 orders of magnitude, e.g., \cite{Weinberg1988CC}. Almost all modified gravity theories do not solve this `old' cosmological constant problem, i.e., why the cosmological constant does not have the very high value predicted by extrapolation of GR or quantum field estimations (though there has been a spate of recent attempts to resolve this in \cite{Kaloper2014,2018arXiv180500470A,2018arXiv180505937K,2018arXiv180505918L}.) It may be helpful though if modified gravity is found responsible for cosmic acceleration, as in that case one can think about the cosmological constant problem with no connection to cosmic acceleration. However, some modified gravity models such as Horndeski theory (see section \ref{sec:horndeski}) or non-local models (see section \ref{sec:nonlocal}) can solve aspects of the `new' cosmological constant problem, that is: ``why is the cosmological constant small, but \textit{not exactly zero}?'' (zero might be reasoned as a natural value if protected by some unknown symmetry). 
An alternative statement of this is: ``why does the cosmological constant have the specific non-zero value measured from observations?'' \citep{Carroll2001CC,Peebles2003,2007-Ishak-Remarks-DE}. This formulation of the problem highlights the coincidental fact that the observed cosmological constant energy density is close to the matter density today. 

It is also important to plan for a future DE/MG analysis now, to ensure that GR/$\Lambda$CDM assumptions are not hidden in analysis pipelines, and that appropriate resources are available for the task when the analysis is eventually performed. If multiple large MG N-body simulations are needed to support the analysis (as is currently the assumption in the Euclid collaboration \citep{Euclid2018}), it is best to be aware of this well in advance to allow time for resource allocation, software and skills development, and the possibility of sharing the workload with other survey collaborations. Finally, and as mentioned earlier, the development of analysis tools to test models beyond $w(z)$CDM and their phenomenological treatments will also enlarge the eventual fundamental cosmology science return from LSST, and will serve as a foundation for developing more sophisticated analyses that may be required by the early 2020s.

\subsection{Overview and selection criteria}
\label{sec:overview}

This document summarizes the results of a consultation process within the TJP working group to identify a set of models that are: (a) physically self-consistent and well-motivated enough to be worth testing; (b) theoretically mature enough that accurate predictions can be made for LSST observables; (c) interesting and different enough from GR/$\Lambda$CDM to serve as good ``templates'' for a broader selection of alternative theories that we may want to test in future; and (d) accessible and meaningfully testable with specifically LSST data, either on its own, or in combination with contemporaneous experiments.

The relative weight of each of these criteria will differ from person to person, and aspects such as ``physical motivation'' are also subjective. These issues were explored and thoroughly discussed at DESC hack weeks, collaboration meetings, and several TJP teleconferences, and by the end of the process we were able to reach a consensus on the recommendations detailed below.

\subsection{Intended uses of this document}
\label{sec:uses}

The recommendations made in this document are intended to provide guidelines to DESC members on which DE/MG theories appear to have higher priority for collaboration efforts due to their perceived greater informative value. This will be useful to the DESC members when performing activities such as building analysis pipelines and planning N-body simulations. The list of recommended models should not be taken as exclusive, nor should it hamper work on other models. Instead, it establishes a ``preferred'' set of reference or example models that DESC members can focus their development work on, in the knowledge that other members should also be prioritizing the same models.

In summary, the recommended models have been selected because they were found to offer a good combination of physical motivation and interest, maturity, and testability with LSST.

\section{Recommendations}
\label{sec:recommendations}

We recommend that development is initially focused on the following class of models:

\begin{enumerate}
 \item {\bf Horndeski theories}, the most general class of scalar-tensor theories that introduce a single scalar field and have at most second-order equations of motion (see also discussion about beyond-Horndeski models further below). These can be parameterized through the ``alpha'' parameterization on linear scales, where their behavior is well-understood theoretically. This provides a nested model that covers all Horndeski theories. Initial constraints exist on some of the alpha parameters, and necessary numerical tools (e.g., Boltzmann solvers) exist and are relatively mature.
\end{enumerate}
We also have a secondary set of recommendations, which can be thought of as a list of the next-highest priority models in line for further study:
\begin{enumerate}
 \item {\bf Bigravity theories}, where a second tensor gravitational field is introduced. These models are closely related to massive gravity theories, where the graviton is allowed a non-zero mass. Bigravity models have a theoretically-compelling internal structure. In some of these models, the cosmological constant does not need to be fine-tuned to an extremely small value as in GR.
 \item {\bf Non-local gravity theories}, that include models with some interesting physical motivations based upon developing a gravitational equivalent to quantum effective actions. In quantum effective actions, light degrees of freedom not appearing in external propagators are integrated out, resulting in an effective action with non-local terms \citep{2016arXiv160608784M}. Non-local gravity models have also been shown to have a viable Friedmann-Lemaitre-Robertson-Walker (FLRW) background evolution, stable linear perturbations, and be a good fit to  current observations. Full LSS and CMB analysis tools are available and N-body simulations are under development by other groups. However, very recent works have shown that some of the models passing current cosmological tests do not pass tests from Lunar Laser Ranging, see e.g., \citep{Belgacem2018LLR}.  
 \item {\bf General $f(R)$ theories}. While cosmological $f(R)$ models are essentially ruled out by astrophysical constraints, these remain useful for studies of  observable effects on intermediate (galactic/extragalactic) scales and can serve for testing pipelines and frameworks.

\end{enumerate}

It is worth mentioning for clarification, that our recommendations above are for specific MG models or classes of models. It is understood that LSST will constrain MG parameters using the Poisson-slip approaches. Parameterized frameworks for modified gravity aim to embody the features of an entire family of models in terms of a few parameters or functions. Horndeski gravity, our top recommendation above, is effectively a parameterization of scalar-tensor gravity theories. A more phenomenological parameterization can be developed by allowing multiplicative modifications to the gravitational Poisson equation and weak lensing potential. This approach has been widely used to date; its simplicity makes it a useful `first step' in any model-independent analysis. For completeness, we will discuss this approach in section \ref{sec:parameterizations}. 
 
Finally, we make separate recommendations on the models that should be used for N-body simulations, based on practical considerations such as the availability of simulation codes and expertise (more than how compelling the theories are).

\subsection{Horndeski theories}
\label{sec:horndeski}

\paragraph{Overview.} The Horndeski class of models represents the most general modification of GR that features a single additional scalar field with no more than second-order derivatives in its equation of motion \citep{Horndeski:1974wa,Deffayet:2011gz}. The Horndeski family contains a number of well-known models as sub-classes, including quintessence, Brans-Dicke, Galileons, and $f(R)$ theories \citep{2011PhLB..706..123D}, with yet more models falling into the extended beyond-Horndeski family, discussed further below. Specific models within the Horndeski class correspond to choices of four functions of a scalar field and its kinetic term; these functions act as coefficients of terms in the action, and are often written as `$G_i$' functions, see Eq.~\ref{eq:HDlagrangian} below. At the linearized level, these functions map onto an alternative set of four functions of time (equivalently, redshift) that describe certain physical properties of the theory; this is the `$\alpha$' parametrization of \cite{Bellini14}. 

The Horndeski class is well-explored theoretically: considerations of stability and the absence of ghosts have been used to derive viability bounds in the space of alpha parameters \citep{Kennedy2018,Denissenya2018}. Three Boltzmann codes exist for calculating observables: EFTCAMB, hi\_CLASS and COOP \citep{Zumalacarregui:2016pph,EFTCAMB1,EFTCAMB2,2016PhRvD..93d3538H}. hi\_class is formatted natively in terms of the $\alpha$ parameters; EFTCAMB operates in an alternative notation used in developing the Effective Field Theory (EFT) approach to dark energy and modified gravity, but possesses a module to map these to the familiar $\alpha$s; COOP likewise possesses functionality for both the $\alpha$ and EFT notation. A comparison of these Einstein-Boltzmann codes can be found in \cite{2018PhRvD..97b3520B}. There have also been some recent observational tests of the Horndeski class \citep{Bellini:2016kl,Kreisch:2017ws}, and further theoretical work on their conditions for viability and range of validity (see below).

In this section, we give an overview of the Horndeski class of models and the current constraints on these models. We begin by giving the motivation for the Horndeski models and some brief notes on extensions of the original Horndeski family. We then describe the current observational constraints on these theories, and recent theoretical developments which can potentially narrow down the space of viable models.  The final sections contain information on the main equations governing these models and how these translate to linear perturbation theory.

\subsubsection{Motivation and extensions}
\paragraph{Motivation.} 
Lovelock's theorem gives a short list of conditions that GR, uniquely, satisfies \citep{Lovelock1971, 2012-Clifton-MG}. One can classify MG theories according to which of these conditions they break. One of these conditions is that the only active gravitational degrees of freedom are the two spin-2 excitations of the GR metric. Horndeski theories are the most general class of theories that break this condition by adding a single extra scalar field (and therefore an extra gravitational degree of freedom), whilst maintaining second-order equations of motion, and without breaking other conditions (e.g., allowing higher-order derivatives, extra dimensions, or a non-local action). The majority of `dark energy' models in the literature fall within the Horndeski class, making them the most broadly studied class of theories. 

Certain subclasses of Horndeski theories have compelling properties, such as the ability to self-accelerate and the presence of screening mechanisms that allow them to pass Solar System tests. Arguably, almost all self-accelerating Horndeski models need finely-tuned parameter values in order to explain cosmic acceleration (see section \ref{sec:intro}). However, this does not necessarily deprecate their value. The more realistic hope is that, were a modified gravity theory to be shown to be observationally preferred, the new developments in physics it triggers would (in time) explain these parameter values in a more natural way. For the present, modified gravity theories such as the Horndeski family can be considered as toy models exhibiting a range of  gravitational phenomena that cosmology experiments like LSST can probe. 

\paragraph{Extended models.}
It was shown in some recent studies \citep{Zumalacarregui:2013pma, Gleyzes15a,Gleyzes15b,Crisostomi16,Crisostomi17} that single scalar-tensor theories can be extended beyond Horndeski models. These extended models have equations of motion that have higher order derivatives, {\color{black}but due to internal constraints they remain immune from Ostrogradski instabilities (modes that decay to infinitely negative energies, a generic pathology of higher-derivative theories \citep{2007LNP...720..403W})} . In fact, this subclass of Lagrangians, dubbed `Beyond Horndeski', can be mapped onto regular Horndeski Lagrangians by a disformal transformation\footnote{\color{black}A generalization of a conformal transformation involving a scalar field, written schematically as
 \begin{equation} {\bar{g}_{\mu\nu}=C(\phi, X)g_{\mu\nu}+D(\phi,X)\nabla_\mu\phi\nabla_\nu\phi}
 \end{equation}
  where $C$ and $D$ are arbitrary functions of the scalar field $\phi$ and its kinetic term $X=-\nabla_mu\phi\nabla^\mu\phi/2$. Regular conformal transformations are the special case $D=0$, $C=C(\phi)$.}-- hence it \textit{must} be the case that they retain only second-order equations of motion. The Beyond Horndeski extensions are described by an additional two $G_i$ functions at the action level, though these collapse to only one additional $\alpha$ parameter ($\alpha_H$, see \S\ref{sec:horndeski:implementation}) at the linearized level.

Further study of the structures leading to degeneracy conditions allowed \cite{Langlois16,Langlois16b} to find a yet more general class of models called Derivative Higher Order Scalar-Tensor (DHOST). Again, due to the degeneracy relations in their Lagrangian, DHOST models are ghost-free despite having higher order derivative equations of motion. DHOST models have been discussed and classified further in \cite{Crisostomi16,Crisostomi17,BenAchour16,BenAchour16b}. When applied to a cosmological spacetime, only one stable DHOST model remains, which is equivalent to Beyond Horndeski under a conformal transformation.

Beyond Horndeski and DHOST models are subject to the constraints from gravitational waves in a manner similar to Horndeski gravity (see our discussion in \S\ref{sec:horndeski:constraints:gravwaves}). However, due to their larger number of $G_i$ functions they do offer a larger spectrum of models that pass both the GW170817/GRB170817A constraint on gravitational wave speed, and some astrophysical constraints \citep{Crisostomi17}. Some of these models are able to drive an accelerating universe \citep{Crisostomi17c}, although recent work suggests they are generically ruled out by an instability to graviton decay (see below). 

Hence it seems likely that most Beyond Horndeski/DHOST models are non-viable as cosmological gravity models; the special cases that remain viable share much of the basic phenomenology of regular Horndeski gravity. Hence, for the present, we will focus our efforts on testing the standard Horndeski family, for which computational and theoretical tools are better-developed. We flag these extensions of Horndeski as interesting avenues to pursue at a later stage of LSST, if possible.

\subsubsection{Observational effects and status}

Horndeski theories generically modify the background expansion (parametrized by the effective equation of state of dark energy, $w(z)$), the growth rate of matter perturbations, the lensing potential, and the propagation speed of gravitational waves. There are also significant effects on non-linear scales; in some members of the Horndeski class, there may exist a `screened' region (i.e., where deviations from GR are suppressed) on small scales. We here describe how Horndeski models modify the background evolution and evolution of perturbations in the universe. We also summarize the constraints placed on these theories by gravitational waves. As we explain, there are still interesting regions of parameter space that LSST will probe.

\paragraph{Background.} The range of functional forms for $w(z)$ in Horndeski gravity is determined by the $G_i$ Lagrangian functions, as shown in Eqs.~\ref{eq:Friedmann}-\ref{eq:pressure} below. In principle, these same $G_i$ functions control the perturbative dynamics of the theory. However, the $\alpha$-parameterization of Horndeski described above separates out the behaviour of linear perturbations from the background expansion. Operationally this is convenient, since it allows one to more or less arbitrarily choose a function (such as $H(z)$ or $w(z)$) that describes the background expansion rate, and constrain the $\alpha_i$ separately from this. Some specific subclasses of Horndeski may have stronger restrictions on $w(z)$, e.g., quintessence models, which have $w \ge -1$. 

While modifications in the perturbative sector do have some dependence on the choice of $w(z)$, most effects survive even if one sets $w=-1$. In other words, in the $\alpha$-parameterization, one can have a background expansion that is fully consistent with $\Lambda$CDM, but growth rate and lensing potential predictions that are different. As such, current constraints on $w(z)$ are not particularly constraining for Horndeski models. However, if in the future we find $w \neq -1$, this can rule out certain subclasses of Horndeski. 

\paragraph{Subhorizon linear scales.} The effects of Horndeski theories on matter perturbations can be somewhat more complicated. If perturbations of the scalar field are sufficiently slowly-varying (i.e., varying on time scales of order the Hubble rate, $H$), and one concentrates only on sub-horizon distance scales (the quasi-static limit), then relatively simple expressions can be found for the effects on growth and lensing. These are often parametrized as effective Newton's constants for these two phenomena, $G_{\rm matter}$ and $G_{\rm light}$, or related parametrizations (e.g., the $\mu$ and $\Sigma$ parameters) at the level of the Poisson and lensing equations. Checks of the quasistatic approximation have been performed for several theories \citep{Noller2014,Sawicki2015,Bose2015,Burrage2017, Lagos2017}; thus far, it has always been found to be an excellent approximation on the scales of current galaxy surveys (with the possible exception of void interiors \citep{Barreira2015,Winther:2015pta,2018arXiv180307533B}).

In the quasi-static limit, the $G_{\rm matter}$ and $G_{\rm light}$ corresponding to Horndeski theories take the form of a ratio of polynomials of wavenumber, multiplied by functions of time \citep{2013PhRvD..87b3501A,2013PhRvD..87j4015S}. However, forecast studies have indicated that the scale-dependence of $G_{\rm matter}$ and $G_{\rm light}$ is generically difficult to detect, unless the mass of the scalar field is very different from the Hubble scale (see, e.g., \citealt{Baker14}). Hence, under moderate assumptions, it is common to approximate $G_{\rm matter}(a)$ and $G_{\rm light}(a)$ as pure functions of  time redshift; their relation to the Horndeski $\alpha$ parameters is discussed below. Various combinations of these functions have been constrained observationally, typically with strong assumptions about the functional forms of the parameters \citep{2010PhRvD..81j3510Z,2010PhRvD..82j3523D,2013MNRAS.429.2249S, Joudaki2017, Ferte2017, DESY1ext}. 
We do not see these constraints as binding for the general Horndeski class, since they are made using these very particular choices of functional form.

\paragraph{Ultra-large scales.} Some modifications allowed within the Horndeski class are only testable on those very large scales on which the quasi-static approximation is not valid (specifically, when studying scales comparable to or larger than the sound horizon of the scalar \cite{2015PhRvD..92h4061S}). For example, non-zero values of the Horndeski kineticity parameter, $\alpha_K$, typically only have effects outside the quasi-static limit. Boltzmann codes that solve the full system of linear evolution equations exist (e.g., the aforementioned hi\_CLASS, EFTCAMB, and COOP) and can calculate predictions for observables outside the quasi-static limit. The full linear theory expressions are given in Appendix \ref{app:horndeski}.

\paragraph{Non-linear scales.} Several subclasses of Horndeski theories exhibit screening mechanisms, which are strongly non-linear effects that cause beyond-GR phenomena to be strongly suppressed in certain environments, generally corresponding to regions of high density. This allows these theories to evade Solar System and binary pulsar constraints on GR, which are otherwise extremely tight. Several different screening mechanisms are known -- Chameleon screening, Symmetron screening, the Dilaton mechanism, K-mouflage and Vainshtein screening. Strong astrophysical constraints exist on models that exhibit Chameleon and Symmetron screening \citep{2017-Burrage-Sakstein}, and so most of these models are essentially excluded observationally. This includes the popular Hu and Sawicki $f(R)$ model.

Note that the relationship between Horndeski theories on large scales and screening on small scales is not always clear in the general case (i.e., outside specific, well-defined subclasses). Constraints on small scales cannot always be clearly translated to constraints on the parameterizations. There are also serious unresolved conceptual issues surrounding Vainshtein screening. Its highly non-linear nature has made calculations very difficult except in a small number of idealized cases; thus, it is unclear how well the mechanism works in dynamical situations and non-spherical matter distributions. It is known to break down (i.e., allowing deviations from GR to manifest) in the interior of some matter sources \citep{2018PhRvD..97b1301C,PhysRevLett.116.061101}. 

Nevertheless, a number of N-body codes exist that make specific predictions for phenomena on quasi-linear and non-linear cosmological scales, including effects due to screening. To date, these require a specific model to be specified from within a given subclass of Horndeski. However, simulations have been performed for a significant number of examples of such models\footnote{Simulations we are aware of are: $f(R)$ gravity (several models), cubic and quartic galileons, DGP and nDGP, symmetron (several models), chameleon (several models), dilaton (several models), non-local gravity, Brans-Dicke gravity, disformal scalar-tensor theories, and coupled quintessence.}. This has given us a good idea of what sort of phenomena we could expect on mildly non-linear scales for some of these theories. It has not yet been possible to build a simulation code that is able to make predictions for any model within the general Horndeski class (see section \ref{sec:nlsims}). Thus, simulation efforts remain subclass-dependent at present. These model-specific simulations have been used to interpret observations of astrophysical systems, such as dwarf galaxies and cluster dynamics, which have put meaningful constraints on some subclasses of Horndeski theories; see \cite{Winther:2015wla} for a comparison of the main model-specific codes.

\paragraph{Gravitational waves.}\label{sec:horndeski:constraints:gravwaves} The recent simultaneous detection of gravitational waves (hereafter GWs) and electromagnetic radiation from a binary neutron star merger has put extremely stringent constraints on the propagation speed of gravitational waves, $c_T$ \citep{Lombriser2016,Lombriser2017C,Baker2017,Zuma2017,Creminelli2017,2017PhRvL.119y1303S,2018arXiv180709241M}. Since one of the Horndeski $\alpha$ parameters controls GW speed as $c_T^2 = 1 + \alpha_T$, these events have major implications for sectors of the Horndeski family. Essentially, any subclass that predicts $\alpha_T \neq 0$ is now strongly constrained. At the level of the $G_i$ Lagrangian functions, the standard interpretation is that $G_5=0$ and $G_4$ can only be a function of $\phi$ (i.e., a conformal factor multiplying the Einstein-Hilbert term). There are a few exceptional `loophole' cases that evade these constraints, but they are generally considered to be very fine-tuned (see \cite{Baker2017} for details).

Some popular Horndeski subclasses -- such as Gauss-Bonnet gravity and the quartic and quintic Galileons\footnote{Note that the cubic Galileon, though consistent with GW observations, is in strong tension with other observables \citep{Renk2017}.} -- are therefore ruled out, unless the above fine-tuning is performed. Other subclasses -- including quintessence, Brans-Dicke, and Kinetic Gravity Braiding models -- predict $\alpha_T=0$ anyway, and so are not constrained by the gravitational wave observations. While several of the surviving model classes are in some sense `minimal' modifications to GR (e.g., quintessence is just a minimally-coupled scalar field, with no couplings to the matter sector beyond the standard gravitational interaction), models with interesting phenomenology do remain viable. These include models with non-universal couplings to baryons and dark matter, models with very slowly evolving scalar fields (which naturally leads to a suppressed value of $\alpha_T$, \citep{Baker2017}), models in which the sound horizon is less than the Hubble scale such as $k$-essence, and models with intrinsic mass scales much larger than the Hubble scale\footnote{One can achieve $\alpha_T\ll 1$ by making $\dot{\phi}^2\ll m^2$, where $m$ is a characteristic mass scale introduced by the MG model (typically $m\sim H_0$) and $\phi$ has been normalized so that it is dimensionless. Hence the equality is satisfied by either very slowly-evolving fields and/or $m\gg H_0$ (see \url{https://www.novelprobes.org/living-review-1} for further discussion).}. Not all of these surviving models are able to accelerate the background expansion rate without a cosmological constant. Note, however, that surviving, self-accelerating models do exist within the Beyond Horndeski/DHOST extension (with some fine-tuning) \citep{Crisostomi17c}.

\paragraph{Theoretical developments in the aftermath of GW170817}
The events surrounding the binary neutron star merger sparked a wave of wider investigations into the behaviour of gravitational waves in modified gravity theories, and in particular in Beyond Horndeski theories and DHOST. Aside from the primary constraint on $c_T$ described above, we describe here two more subtle realisations that have emerged (note that this is not an exhaustive list):\newline

 {\bf i) Cut-off scales for Horndeski.} It is important to remember that Horndeski gravity and its extensions are low-energy effective theories, constructed to affect cosmological scales. This means that at some higher energy scale we should expect new operators, not already present in the Horndeski Lagrangian, to become relevant, and cause our EFT to break down. The authors of \cite{deRhamGW2018} have pointed out that the frequency of GW170817 at merger ($\sim$ 100 Hz) lies close to the energy scale where this might be expected to occur for Horndeski gravity. Furthermore, if Horndeski gravity is to be the low-energy limit of a Lorentz-invariant UV-complete theory, it \textit{must} acquire new operators that cause it to asymptote as $c_T\rightarrow 1$ at high energies.

 It is therefore possible for a Horndeski model which superficially appears ruled out (according to the arguments of the previous subsection), to actually be compatible with the observations of GW170817, whilst substantially affecting cosmological scales. Ultimately, the relevant cut-off scale depends on the specifics of a given model, so we cannot make general statements here; a concrete example is provided in \cite{deRhamGW2018}. Similar conclusions were reached through a different line of reasoning in \cite{BattyeGW2018}.

{\bf ii) Graviton Decay.} Recent work by \cite{Creminelli2018} studied scenarios in which a graviton ($\gamma$) is able to decay into fluctuations of the Horndeski scalar field ($\pi$), via the channels $\gamma\rightarrow \pi\pi$ and/or $\gamma\rightarrow \pi\gamma$. The authors found that the necessary operators exist (and indeed, the decay rates are generically large) in theories involving the Horndeski $G_{4,X}$ and $G_5$ derivative operators, as well as the Beyond Horndeski $F_4$ and $F_5$ operators. The former two operators are already constrained (subject to the caveat of i) above) by the bounds on $c_T$. However, this result further implies that virtually all of the Beyond Horndeski parameter space spanned by $\{F_4, F_5\}$ is also ruled out, by the simple fact that GWs reach us at all without completely decaying to $\pi$. One exception to this argument is theories which have the sound speed of the scalar field equal to unity, $c_S^2=1$; for these models no decay occurs. At present, the relation between this graviton-based calculation and classical gravitational waves is yet to be fleshed out. Assuming the conclusions remain unchanged for classical waves, these results could potentially eliminate {\color{black}the majority of extended Horndeski models}.

\subsubsection{Action and background equations}
This section gives a short summary of the equations describing the Horndeski class of models. This will serve as a reference for the interested reader. Note that the parameterizations we recommend implementing are described in later sections.

Full linear theory expressions exist for the Horndeski class in its full generality. Two key resources for linear perturbation theory in Horndeski are \cite{Bellini14} and \cite{Gleyzes14}. The former was the first work to present the Horndeski $\alpha$ parameters (see below) (notation which was later implemented in the hi\_Class Boltzmann code). The latter contains an extension to a fifth `beyond Horndeski' parameter, $\alpha_H$, and an extension to non-pressureless matter.

In the notation of  \cite{Bellini14}, the Lagrangian for Horndeski gravity is
\begin{eqnarray}
S&=&\int\mathrm{d}^{4}x\,\sqrt{-g}\left[\sum_{i=2}^{5}{\cal L}_{i}\,+\mathcal{L}_{\text{m}}[g_{\mu\nu}]\right]\,,\label{eq:HDlagrangian}\\
{\cal L}_{2} & = & K(\phi,\, X)\,,\nonumber\\
{\cal L}_{3} & = & -G_{3}(\phi,\, X)\Box\phi\,,\nonumber\\
{\cal L}_{4} & = & G_{4}(\phi,\, X)R+G_{4X}(\phi,\, X)\left[\left(\Box\phi\right)^{2}-\phi_{;\mu\nu}\phi^{;\mu\nu}\right]\,,\nonumber\\
{\cal L}_{5} & = & G_{5}(\phi,\, X)G_{\mu\nu}\phi^{;\mu\nu}-\frac{1}{6}G_{5X}(\phi,\, X)\left[\left(\Box\phi\right)^{3}+2{\phi_{;\mu}}^{\nu}{\phi_{;\nu}}^{\alpha}{\phi_{;\alpha}}^{\mu}-3\phi_{;\mu\nu}\phi^{;\mu\nu}\Box\phi\right]\,,\nonumber
\end{eqnarray}
where $X \equiv -\frac{1}{2}\nabla^\mu\phi\nabla_\mu\phi$, and $G_{4X} \equiv dG_4 / dX$. We see that there are four $G_i$ functions in this action; the function $K$ is alternatively named $G_2$ by some authors ($K$ stands for `kinetic', since this part of the Lagrangian contains the standard kinetic term of the scalar field). The $G_i$ quantify the amplitude of successively higher-derivative contributions to the Horndeski Langrangian; yet, the structure of the Lagrangians is such that the resulting field equations contain at most second-order derivatives.

The Friedmann equations are obtained from this action in the standard way. Unusually, though, one finds that the Planck mass appearing in these equations is replaced by an effective Planck mass that potentially evolves with time. This effective Planck mass is given by:
\begin{equation}
M_{*}^{2}\equiv2\left(G_{4}-2XG_{4X}+XG_{5\phi}-\dot{\phi}HXG_{5X}\right)\,,\label{eq:MPl}
\end{equation}
where $\phi$ here is the homogeneous value of the scalar field (i.e., $\phi=\bar\phi$). From the above expression one defines the parameter $\alpha_M$ as the effective Planck mass run rate:
\begin{equation}
\alpha_{\textrm{M}}\equiv H^{-1}\frac{\mathrm{d}\ln M_{*}^{2}}{\mathrm{d}t}\,.\label{eq:omega1}
\end{equation}
This $\alpha$ parameter is one of a set that we will introduce shortly. The Friedmann equations containing $M_*^2$ are (where dots denote derivatives with respect to physical time):
\begin{align}
 & 3H^{2}=\frac{1}{M_*^2}\left[{\rho}_{\text{m}}+{\mathcal{E}}\right]\label{eq:Friedmann}\\
 & 2\dot{H}+3H^{2}=-\frac{1}{M_*^2}\left[{p}_{\text{m}}+{\mathcal{P}}\right]\nonumber 
\end{align}
{Here $\rho_m$ and $p_m$ are respectively the energy density and pressure of matter}, and the effective energy density and pressure of the Horndeski sector are:
\begin{eqnarray}
{\mathcal{E}} & \equiv & -K+2X\left(K_{X}-G_{3\phi}\right)+6\dot{\phi}H\left(XG_{3X}-G_{4\phi}-2XG_{4\phi X}\right)\label{eq:density}\\
 &  & +12H^{2}X\left(G_{4X}+2XG_{4XX}-G_{5\phi}-XG_{5\phi X}\right)+4\dot{\phi}H^{3}X\left(G_{5X}+XG_{5XX}\right)\,,\nonumber \\
{\mathcal{P}} & = & K-2X\left(G_{3\phi}-2G_{4\phi\phi}\right)+4\dot{\phi}H\left(G_{4\phi}-2XG_{4\phi X}+XG_{5\phi\phi}\right)\label{eq:pressure}\\
 &  & -M_{*}^{2}\alpha_{\textrm{B}}H\frac{\ddot{\phi}}{\dot{\phi}}-4H^{2}X^{2}G_{5\phi X}+2\dot{\phi}H^{3}XG_{5X}\,.\nonumber 
\end{eqnarray}
Note that the expression for $\cal P$ features one further $\alpha$ parameter, $\alpha_\textrm{B}$ (see below). As we discussed above, a common approach is to simply parameterize the dark energy equation of state in the usual way, $w_X = {\cal P}/{\cal E}$; in this case the detailed form of the above expressions is not needed. 

Note that $\rho_m$ obeys its usual conservation law, but the quantity $(\rho_m/M_*^2)$ does not, due to the time-dependence of the denominator. Some Einstein-Boltzmann solvers are natively coded to evolve $(\rho_m/M_*^2)$, so some care is needed. Either the code must be rewritten to operate in terms of $\rho_m$, or a modified matter conservation equation must be used (see eqs.3.9 of \cite{Bellini14}).

\subsubsection{Linear perturbations}
\label{sec:horndeski:implementation}
Here we describe the main parameterizations that we recommend for implementation.

\paragraph{Alpha parametrization.} 
We will take our definition of the perturbed line element in the conformal Newtonian gauge to be:
\begin{align}
\label{lineel}
ds^2&=-\left(1+2\Psi\right)dt^2+a(t)^2\left(1-2\Phi\right)dx_idx^i
\end{align}
The full set of linearized field equations for Horndeski gravity are given in the Appendix. They feature five important quantities identified as the $\alpha$ parameters. These $\alpha$ parameters are essentially linear combinations of the $G_i$ functions that appear `natively' in the Horndeski action, plus their derivatives with respect to $\phi$ and its kinetic term (and factors of $H$, $\dot\phi$, etc.). They were selected as useful quantities to work with because they are easier to interpret in terms of physical effects; in some sense, they are more closely connected to observables than the original $G_i$. 

We have met the Planck mass run rate parameter $\alpha_\textrm{M}$ already, in eqs.(\ref{eq:MPl}) and (\ref{eq:omega1}). The remaining $\alpha$s are:
\begin{itemize}
\item {\bf $\alpha_\textrm{B}$, the braiding parameter}: non-zero values of this parameter create an interaction between the kinetic terms of the scalar field and the metric. This results in second derivatives of the metric perturbation appearing in the equation of motion for the scalar field, and vice-versa. It also potentially introduces a `braiding scale' into the gravitational field equations, where the dynamics of field perturbations undergo a qualitative change. This does not necessarily lead to scale-dependent behaviour in observables, though, if the braiding scale is very close to the cosmological horizon.  
\item  {\bf $\alpha_\textrm{K}$, the kineticity parameter}: {\color{black} A canonical scalar field has kinetic term $X$, so the more general function $K(\phi,X)$ can be thought of as a generalization of the kinetic energy of the scalar field perturbations.} For a quintessence model with $w \ne -1$, this is the only non-zero $\alpha$ parameter. Large values of $\alpha_\textrm{K}$ suppress the sound speed of the scalar perturbations. Notably, $\alpha_\textrm{K}$ only impacts ultra-large scale modes of cosmological perturbation theory; it drops out of the relevant quasi-static expressions. For this reason, analyses to date have marginalized over or fixed $\alpha_\textrm{K}$, rather than attempt to constrain it.
\item  {\bf $\alpha_\textrm{T}$, the tensor speed excess parameter}: non-zero values of this parameter indicate that tensor perturbations of the metric (gravitational waves) propagate at a speed different from the GR prediction of $c$. Their speed is instead given by $c_T^2 = c^2 (1+\alpha_\textrm{T})$, where $\alpha_\textrm{T}$ can be either positive or negative\footnote{Negative values of $\alpha_\textrm{T}$ are also constrained \textit{for some energy ranges} by a lack of observed gravi-Cherenkov radiation \cite{2001JHEP...09..023M}.}. Non-zero $\alpha_\textrm{T}$ also contributes to effective anisotropic stress, causing the metric potentials $\Phi$ and $\Psi$ to differ. As described above, the binary neutron star merger event GW170817 and its electromagnetic counterparts have ascertained the bound $|\alpha_\textrm{T}|\lesssim 10^{-12}$ (the bound can be tightened by three orders of magnitude under less conservative assumptions). 
\item There is also a fifth $\alpha$ parameter, the {\bf `Beyond Horndeski' parameter $\alpha_\textrm{H}$}. Non-zero values of $\alpha_\textrm{H}$ indicate that there exist equivalent Lagrangians for a single theory, which are related to each other by disformal transformations (the generalization of a conformal transformation -- see the introduction of this section for a brief explanation). {\color{black} In these equivalent theories, the derivatives $G_{4X}$ and $G_{5X}$ that appear in the fourth and fifth lines of eq.\ref{eq:HDlagrangian} are replaced by more general functions. In other words, the relationship between the two terms in each of $L_4$ and $L_5$ is broken. In the case of $\alpha_H=0$, the form of eq.\ref{eq:HDlagrangian} is recovered.}

The relation of $\alpha_\textrm{H}$ to the $G_i$ functions is more complex than those given below; see \cite{Gleyzes14} for full details. We recommend that the initial LSST constraints on modified gravity hold the $\alpha_\textrm{H}$ parameter fixed to zero, as i) cases with $\alpha_\textrm{H}\neq 0$ are not supported in the current public release of the hi\_Class code, ii) the principal known effects of $\alpha_\textrm{H}$ are on stellar structure, and not cosmological-scale observables \citep{2015PhRvL.115t1101S,2015PhRvD..92l4045S}, and iii) the recent results of \cite{Creminelli2018} suggest that most theories with $\alpha_H \neq 0$ are disfavored by graviton decay.
\end{itemize}
We anticipate that LSST tests of Horndeski gravity will focus on constraining the first four $\alpha$ parameters, as this is the framework implemented in the hi\_Class code. However, it is often useful to relate the $\alpha$s to the native $G_i$ functions, in order to understand consequent implications for the structure of the Horndeski Lagrangian. We give these relations for $\alpha_\textrm{B}$, $\alpha_\textrm{K}$ and $\alpha_\textrm{T}$ here; the equivalent for $\alpha_\textrm{M}$ was given in Eqns.~\ref{eq:MPl} \& \ref{eq:omega1} above.
\begin{align}
H^{2}M_{*}^{2}\alpha_{\textrm{K}}&=  2X\left(K_{X}+2XK_{XX}-2G_{3\phi}-2XG_{3\phi X}\right)+\label{eq:omega2}\\
& +12\dot{\phi}XH\left(G_{3X}+XG_{3XX}-3G_{4\phi X}-2XG_{4\phi XX}\right)+\nonumber \\
& +12XH^{2}\left(G_{4X}+8XG_{4XX}+4X^{2}G_{4XXX}\right)-\nonumber \\
& -12XH^{2}\left(G_{5\phi}+5XG_{5\phi X}+2X^{2}G_{5\phi XX}\right)+\nonumber \\
& +4\dot{\phi}XH^{3}\left(3G_{5X}+7XG_{5XX}+2X^{2}G_{5XXX}\right)\nonumber \\
HM_{*}^{2}\alpha_{\textrm{B}}&=  2\dot{\phi}\left(XG_{3X}-G_{4\phi}-2XG_{4\phi X}\right)+\label{eq:omega3}\\
& +8XH\left(G_{4X}+2XG_{4XX}-G_{5\phi}-XG_{5\phi X}\right)+\nonumber \\
& +2\dot{\phi}XH^{2}\left(3G_{5X}+2XG_{5XX}\right)\nonumber \\
M_{*}^{2}\alpha_{\textrm{T}}&=  2X\left(2G_{4X}-2G_{5\phi}-\left(\ddot{\phi}-\dot{\phi}H\right)G_{5X}\right)\label{eq:omega4}
\end{align}
Note that certain works in the literature may use slightly different defintions of the $\alpha$ parameters (e.g., which differ by a constant factor). As an example use of these expressions, the result from the gravitational wave event GW170817 $\alpha_T\simeq 0$ translates into implications for the Lagrangian terms ${\cal L}_4$ and ${\cal L}_5$ (see, e.g., \citealt{Baker2017}).

\paragraph{Quasi-static limit}
As described above, on linear sub-horizon scales the full modified Einstein equations can be mapped onto a simplified form. This simplified form consists of two functions of time and scale; one of these represents an effective modification to the gravitational Poisson equation:
\begin{align}
\label{paramPoisson}
-2\frac{k^2}{a^2}\Psi&=8\pi G_{\rm eff} (z) \rho_m\delta_m \\
&=8\pi G_N \left(\frac{M_P^2}{M_*^2}\right)\left[1+\mu (z)\right] \rho_m\delta_m \label{bnm}\\
{\rm{i.e.}}\quad\quad\frac{G_{\rm eff}(z)}{G_N}&=\left(\frac{M_P^2}{M_*^2}\right)\left[1+\mu (z)\right]\label{cvb}\\
{\rm{and}}\quad\quad \Psi&=\Psi_{GR}\left(\frac{M_P^2}{M_*^2}\right)[1+ \mu(z)]\label{poi}
\end{align}
Note also the factors of $\left(\frac{M_P^2}{M_*^2}\right)$: these are necessary to account for modifications to the effective Planck mass in the Horndeski Lagrangian.

The second quasi-static function can be chosen as either the ratio of the two metric potentials (often called the gravitational slip) or as a modification to the lensing potential. Since both are in common use, we will show here how the three are related. The slip parameter is defined as $\gamma (z) = {\Phi/\Psi}$, whilst the lensing function $\Sigma(z)$ relates the lensing potential (sometimes called the Weyl potential) to the GR potential by:
\begin{align}
-\frac{k^2}{a^2}\left[\Phi+\Psi\right]&=8\pi G_N\left(\frac{M_P^2}{M_*^2}\right)\left[1+\Sigma (z)\right] \rho_m\delta_m\label{rty}\\
&=-\frac{k^2}{a^2}\Psi_{GR}\left(\frac{M_P^2}{M_*^2}\right)\,2[1+\Sigma(z)]\label{hfd}
\end{align}
in the absence of anisotropic shear. Note the factor 2 above, such that the GR limit is $\Sigma=0$.

The relation between $\Sigma$, $\mu$ and $\gamma$ is found as follows (using eqs.\ref{poi} and \ref{hfd}):
 \begin{align}
\left(\Phi+\Psi\right) &= \Psi\left[1+\gamma (z)\right] \\
&=\Psi_{GR}\,\left(\frac{M_P^2}{M_*^2}\right)\left[1+\mu(z)\right]\left[1+\gamma (z)\right] \\
 \Rightarrow \Sigma (z) &= \frac{1}{2}\left\{\mu(z)\left[1+\gamma (z)\right] +\gamma(z)-1\right\}\label{Sigmadef}
\end{align}
It is important to note that, the way we have defined things here, the ratio $\left(\frac{M_P^2}{M_*^2}\right)$ does not appear in the definitions of $\mu$ and $\Sigma$; however, it must be included in implementations of eqs.(\ref{bnm}) and (\ref{rty}) if $\alpha_M\neq 0$. More details on general model-independent measurements of $\mu$ and $\Sigma$ will be given in Section \ref{sec:musigma}.

Two methods of taking the quasistatic limit have been used in the literature, which only agree in the extreme $k\rightarrow\infty$ limit. It is now generally agreed that the correct method is as follows: combine the full set of linearized field equations to eliminate variables, until one is left only with matter sources and one gravitational degree of freedom (this is usually chosen to be $\Phi$). Set any terms containing time derivatives of the degree of freedom to zero (e.g., $\dot\Phi = \ddot\Phi=0$). The resulting expression relates $\Phi$ to the matter density perturbation, and hence can be rewritten in the form of eq.(\ref{paramPoisson}). \cite{Bellini14} shows some steps of this derivation.


In what follows, we will define the useful quantity:
\begin{align} 
A&= -\frac{\alpha_\textrm{B}}{2}(1+\alpha_\textrm{T})+\alpha_\textrm{T}-\alpha_\textrm{M}
\end{align}
The resulting form for $\mu(z)$ is given in terms of the $\alpha$ parameters by:
\begin{align}
\mu(z)&=\frac{\alpha_T}{2}+\frac{2 A^2}{\DD c_s^2}
 \label{QSPoisson}
\end{align}
where $\alpha$ and $c_s^2$ are: 
\begin{align}
\alpha&=\alpha_\textrm{K}+\frac{3}{2}\alpha_\textrm{B}^2 \label{alpha}\\
c_s^2 &= - \frac{(2-\alpha_\textrm{B}) \Big[ \dot H -  (\alpha_\textrm{M} - \alpha_\textrm{T}) H^2 - H^2 \alpha_\textrm{B}/2 (1+\alpha_\textrm{T})\Big] - H \dot \alpha_\textrm{B} + (\rho_{\rm m} + p_{\rm m})/M_*^2}{H^2 \DD}\label{cs2}
\end{align}
Similarly, the slip parameter obtained via this method is:
\begin{align}
\gamma^{k\rightarrow\infty}(z)= \frac{\DD\,  c_s^2 -\alpha_\textrm{B} A}{\DD\,  c_s^2(1+\alpha_\textrm{T}) + 2A^2}
\label{QSslip}
\end{align}
Note that $\alpha_\textrm{K}$ does not feature in either Eqn.~\ref{QSPoisson} or \ref{QSslip} (it drops out of the combination $\DD c_s^2$), and hence is (effectively) impossible to constrain with data on quasi-static scales. As explained above, $\alpha_\textrm{H}$ has been set to zero in the above expressions. Using the above expressions and Eqn.~\ref{Sigmadef}, $\Sigma^{k\rightarrow\infty}(z)$ can be straightforwardly found as:
\begin{align}
 \Sigma^{k\rightarrow\infty}(z)&=\frac{\alpha_\textrm{T}}{2}+\frac{A\left(A-\alpha_\textrm{B}/2\right)}{\alpha\,c_s^2}\label{yui}
  \end{align}
Current constraints on $G_{\rm eff}$, $\gamma$ and/or $\Sigma$ can be found in (for example) \cite{Planck2015MG,2018MNRAS.474.4894J,2018arXiv181002499D}; see also Section \ref{sec:musigma}.

\subsubsection{Parameters and prior ranges}
\label{sec:horndeski:parameters}

There is no unique, well-defined way of specifying the functional forms that the $\alpha$ functions should take. One could take a series expansion of the $\alpha$ functions, but it is not clear that this is sensible -- as can be seen from Eqs.~\ref{eq:omega2}-\ref{eq:omega4}, the $\alpha$ functions map in a complicated way onto the Lagrangian $G_i$ functions, so choosing a series expansion for one set of functions will correspond to a much more complicated functional form for the other set. It is unclear whether one should prefer the functions in the action ($G_i$-functions), or the functions that parameterize physical properties (the $\alpha$-functions) to be chosen simply.

There are several common choices of functional form for the $\alpha$ parameters in the literature, which do have broadly desirable properties. However, these choices have been broadly motivated by their simplicity, rather than their physical meaning. These are:
\begin{align}
\alpha_i(z) &= \alpha_{i,0}\, \Omega_{\rm DE}(z) \label{DEansatz}\\
\alpha_i(z) &= \alpha_{i,0}\, a^{\gamma_i} \label{PLansatz}
\end{align}
where $\Omega_{\rm DE}(z) = 8\pi G\rho_{\rm DE}(z)/3 H^2(z)$ and $\alpha_{i,0} = \alpha_{i}(z=0)$. The form of Eq.~\ref{DEansatz} has the advantage of decaying to zero at high redshift, meaning that the GR limit is recovered in the early Universe. This is reasonable if one expects modified gravity effects to be purely late-time phenomena. Eqn.~(\ref{PLansatz}) also has this behavior, but the rapidity with which MG effects switch on can be tuned by changing the exponent, $\gamma_i$. It is normal to assume that all of the $\alpha$ functions follow the same functional form, although additional parameters (like $\gamma_i$) might differ from one $\alpha$ function to another. Discussion of the choices of functional ansatz for the $\alpha$ parameters can be found in \cite{2017PhRvD..95b3518L,2017PhRvD..96f3516G}.

Given the difficulty of settling on a functional parametrization for the $\alpha$ functions, it is reasonable to consider `non-parametric' approaches, such as representing the $\alpha$ functions as a series of narrow top-hat bins of unknown amplitude. This is more agnostic in a sense, but has the problem of introducing potentially many more degrees of freedom, which we may find ourselves to be be unable to constrain with any appreciable precision. One should also note that the parametrization issue has not been completely sidestepped here either -- when one comes to constrain the top-hat amplitudes, some choice will need to be made for the prior distribution on these parameters. Since they map nonlinearly to observables, a uniform prior would be informative, and so will potentially bias the results. Without more physically-motivated choices for the functional forms of the $\alpha$ parameters, or at least sensible priors on their smoothness or other properties, we are therefore susceptible to some degree of arbitrariness in our analysis. We expect that community best practice with regards to the choice of functional form of the $\alpha$ parameters will evolve prior to the analysis of LSST data; we recommend the choice of functional form which offers the most information while avoiding bias. Further work is needed to determine the precise optimal choice.


\subsection{Bigravity theories}
\label{sec:bigravity}


\paragraph{Historical overview.} Attempts to build a consistent theory of a massive graviton began almost a century ago, with the seminal work of Pauli and Fierz in the 1930s. They derived the unique, consistent, classical theory of a massive spin-2 field in flat space-time. Crucially, they realized that a particular sign appearing in the mass term of the spin-2 field was key to the consistency of the theory: flipping this sign caused the theory to propagate a ghostly scalar mode. This was the first hint of an eighty-year battle to construct a ghost-free \textit{full} theory (i.e., not just linear) of a massive graviton. 

The Fierz-Pauli Lagrangian lay largely neglected until the 1970s, when van Dam, Veltman and Zakharov realized that, when matter sources are present, the Fierz-Pauli theory does not reduce to GR as the mass of the graviton is taken to zero \citep{VANDAM1970397,1970JETPL..12..312Z}. Instead the zero-mass limit of the theory predicted drastic modifications to gravitational lensing that were already ruled out by observations, even at that time. Given that the Fierz-Pauli Lagrangian was believed to be unique, the lensing catastrophe seemed to rule out any viable theory of a massive spin-2 Lagrangian. The strength of this implication prompted Vainshtein to question the derivation of the so-called `vDVZ discontinuity'. Vainshtein realized that, in fact, a scalar mode of the massive graviton becomes strongly coupled at radii close to a matter source \citep{VAINSHTEIN1972393}. At distances smaller that the \textit{Vainshtein radius}, $r_V$, one needs to calculate using a full nonlinear completion of the theory. This rescued the viability of a massive gravity theory, at least until a complete Lagrangian could be found.

Boulware and Deser tackled the problem immediately, although they returned a negative result. Their claim -- that once again, any nonlinear extension of the Fierz-Pauli Lagrangian was doomed to suffer a ghost \citep{PhysRevD.6.3368} -- stifled work on the theory for the next forty years. Only with the discovery of cosmic acceleration did interest in extensions of GR become strong enough to prompt a reanalysis of the Boulware-Deser ghost.

Initially via a somewhat circuitous route, de Rham, Gabadadze \& Tolley (dRGT \cite{PhysRevLett.106.231101}) were able to construct a full theory of a massive graviton that they showed was ghost-free in a particular limit of the energy scales involved, called the \textit{decoupling limit}. The dRGT massive gravity model employed a second metric that was fixed to be Minkowskian. This was quickly generalised to be an arbitrary, but fixed metric $f_{\mu\nu}$. Hassan \& Rosen finally found a way to give the $f$-metric dynamical behaviour of its own \citep{PhysRevLett.108.041101}, whilst remaining ghost-free: bigravity was born.

\paragraph{Motivation.} Nearly all modified gravity models involve the introduction of new degrees of freedom. We have numerous models that obtain these from spin-0 or spin-1 fields, so it is natural that the spin-2 picture should likewise be explored. Moreover, one could qualitatively argue that spin-2 fields are natural candidates for dark energy, since i) we know that GR itself is formulated in terms of a spin-2 field, hence they are intimately connected with gravity, and ii) we lack a renormalisable quantum theory of spin-2 fields.  The immensely successful Standard Model of particle physics is based on quantum field theories of spin-0, spin-1/2 and spin-1 particles; so any dark energy field from those families would need to preserve existing bounds related to the Standard Model (spin measurements can be inferred from the angular correlations of decay products \citep{1126-6708-2007-05-052}). Spin-2 field theories are less well-explored and hence potentially less subject to current constraints.

Bigravity theories are constructed using a fully dynamical tensor field $f_{\mu\nu}$, which interacts with the regular spacetime metric $g_{\mu\nu}$. The coupling between the two tensor fields is given by a specific set of interaction terms based on elementary symmetric polynomials (see section \ref{subsec:bigravity_action} below). In fact, this is one place where Boulware and Deser went astray in their initial claim that a massive graviton theory could never be ghost-free: they assumed the interaction potentials to lie within a class of functional forms that does not include the elementary symmetric polynomials.

Bigravity propagates seven degrees of freedom (hereafter d.o.f), which is the correct total for a theory of one massless (2 d.o.f.) and one massive (5 d.o.f.) spin-2 fields. Whilst there has been some work on developing Lorentz-violating theories, in this document we adhere to those that preserve Lorentz invariance. The theory can accommodate an isotropic and homogenous expanding background solution (unlike massive gravity with a \textit{fixed} auxiliary metric). Parts of its parameter space can give rise to self-accelerating solutions without a cosmological constant, though these have some caveats that we discuss below. For a review of bigravity theories, see \cite{1751-8121-49-18-183001}.

\subsubsection{Observational effects and status} 
While there is a reasonably extensive theoretical literature on dRGT theory and bigravity, a definitive comparison with the complete cosmological data is still lacking; as yet there has been no calibrated and tested implementation in Einstein-Boltzman solvers which can then be used for parameter estimation. This relative immaturity of the theory must be taken into account when selecting models to priorities for the LSST Beyond-$w(z)$CDM analysis.

As we will explain in the following subsections, we know that it is possible for bigravity models to give rise to self-accelerated cosmologies; however, very large portions of the initially promising parameter space are ruled out by the existence of pathologies (instabilities or ghost modes) for linear perturbations of the model. It has been qualitatively argued that higher-order corrections may turn out to soften these growing modes or cure instabilities, but no concrete proof of this has appeared to date. 

Despite this rather pessimistic status, bigravity retains two particularly appealing features. Firstly, as mentioned above, it was one of few models to be almost unaffected by the observations of the binary neutron star merger GW170817 (with the exception of the sub-case of doubly coupled bigravity, see \cite{PhysRevD.97.124010}). The presence of a massive graviton should cause the group velocity of gravitational waves to differ from $c$, and hence the theory should be subject to constraint from the measured 1.74 seconds delay in the arrival of the GW and EM counterparts of GW170817. In practice the bound obtained in this way  -- $m_g< 10^{-22}$eV -- is not competitive with bounds already in existence. The tightest of these comes from lunar laser ranging experiments, and bounds $m_g<10^{-32}$ eV. A detailed compendium of bounds on the graviton mass can be found in \cite{RevModPhys.89.025004}.

The second appealing feature of bigravity is that it displays Vainshtein screening, similar to the Galileon family of models. This property means that at a radius approaching massive sources -- the Vainshtein radius, $r_v$ -- the interactions of the scalar degree of freedom are suppressed, and hence it should return to a regime similar to GR. Qualitatively, as one nears the matter source, elements of the kinetic matrix of the theory become large; after canonically renormalising, all terms in the Lagrangian are effectively divided by these large values. This means that the coupling of the scalar d.o.f. to matter sources is correspondingly suppressed. Note that the Vainshtein radius is distinct from (and larger than) the cut-off scale, at which the theory becomes unpredictive. Further discussion of the Vainshtein mechanism, in both Galileons and bigravity, can be found in \cite{PhysRevD.88.084002,0264-9381-30-18-184001,2015-rev-Joyce-et-al}.

Although no N-body simulations for bigravity exist, the effects of Vainshtein screening on large-scale structure have been studied through simulations of Galileon gravity and DGP gravity \citep{PhysRevD.80.123003,2015JCAP...12..059B,2013JCAP...11..056B,2015JCAP...12..059B,2018MNRAS.475.3262F,2018arXiv181002864P}. More recently, it has been realised that Vainshtein screening becomes ineffective in time-dependent systems \citep{PhysRevLett.116.061101}, and that this may yield testable predictions for binary systems in some bigravity models \citep{PhysRevD.97.124010}.

\subsubsection{Action and background equations}
\label{subsec:bigravity_action}

In this subsection and the next we present the key background and linearized equations for ghost-free bigravity. The expressions here are taken from \cite{1475-7516-2014-12-026}. The action for massive bigravity is:
\begin{equation}\label{MGaction}
S=\; \frac{M_g^2}{2}\int d^4x \sqrt{-g}R(g) +\frac{M_f^2}{2}\int d^4x \sqrt{-f}R(f) - m^2M_{g}^2\int d^4x \sqrt{-g}\sum_{n=0}^4\beta_n e_n\left(\sqrt{g^{-1}f}\right)+S_m.
\end{equation}
Here $g_{\mu\nu}$ and $f_{\mu\nu}$ are two dynamical metric fields, and $R(g)$ and $R(f)$ are their associated Ricci scalars (kinetic terms). $M_g$ is the regular Planck mass, whilst $M_f$ is the effective Planck mass of the $f$ metric. $m$ is an additional mass scale of the theory. In what follows, we will frequently see the appearance of the ratio $M_*^2\equiv M_f^2/M_g^2$. The action also contains a regular matter action, $S_m$, which couples only to the spacetime metric $g_{\mu\nu}$. 

In addition, this action contains interactions between both metrics that preserve general covariance, and are expressed in terms of the functions $e_n \left(\sqrt{g^{-1}f}\right)$. These correspond to the elementary symmetric polynomials\footnote{\color{black} The elementary symmetric polynomials in $n$ variables $X_1,\ldots,\, X_n$ are denoted $e_k(X_1,\ldots X_n)$ for $k=0,\,1,\ldots,\,n$, and are defined by
\begin{equation}
 e_k(X_1,\ldots X_n) = \sum_{1\leq j_1<\ldots<j_k\leq n}\,X_{j_1}\ldots X_{j_n}
\end{equation} 
 with $e_k(X_1,\ldots X_n) =0$ for $k>n$.} 
 of the eigenvalues $\lambda_n$ of the matrix $\sqrt{g^{-1}f}$, which satisfies $\sqrt{g^{-1}f}\sqrt{g^{-1}f}=g^{\mu\lambda}f_{\lambda\nu}$. The $\beta_n$ are free dimensionless coefficients.

The ensuing equations of motion for $g_{\mu\nu}$ and $f_{\mu\nu}$ are:
\begin{align}
& \; R(g)_{\mu\nu}-\frac{1}{2}g_{\mu\nu}R(g) +\frac{m^2}{2}\sum_{n=0}^3(-1)^n\beta_n\left[g_{\mu\lambda}Y^\lambda_{(n)\nu}\left(\sqrt{g^{-1}f}\right)+ g_{\nu\lambda}Y^\lambda_{(n)\mu}\left(\sqrt{g^{-1}f}\right)\right]=\frac{T_{\mu\nu}}{M_g^2},\label{Eqg}\\
 & \; R(f)_{\mu\nu}-\frac{1}{2}f_{\mu\nu}R(f) + \frac{m^2}{2M_*^2}\sum_{n=0}^3(-1)^n\beta_{4-n}\left[f_{\mu\lambda}Y^\lambda_{(n)\nu}\left(\sqrt{f^{-1}g}\right)+f_{\nu\lambda}Y^\lambda_{(n)\mu}\left(\sqrt{f^{-1}g}\right)\right]=0,\label{Eqf}
\end{align}
where $T^{\mu\nu}$ is the matter stress-energy tensor, and the matrices $Y^\lambda_{(n)\mu}(\mathbb{X})$ are defined as:
\begin{align}
Y_{(0)}=&\mathbb{I},\nonumber\\
Y_{(1)}=&\mathbb{X}-\mathbb{I}[\mathbb{X}],\nonumber\\
Y_{(2)}=&\mathbb{X}^2-\mathbb{X}[\mathbb{X}]+\frac{1}{2}\mathbb{I}\left([\mathbb{X}]^2-[\mathbb{X}^2]\right),\nonumber\\
Y_{(3)}=&\mathbb{X}^3-\mathbb{X}^2[\mathbb{X}]+\frac{1}{2}\mathbb{X}\left([\mathbb{X}]^2-[\mathbb{X}^2]\right) -\frac{1}{6}\mathbb{I}\left([\mathbb{X}]^3-3[\mathbb{X}][\mathbb{X}^2]+2[\mathbb{X}^3]\right),
\end{align}
where $\mathbb{I}$ is the identity matrix and $[\mathbb{X}]$ stands for the trace of the matrix $\mathbb{X}$. We also have local conservation of energy-momentum: 
$\nabla^{\mu}_gT_{\mu\nu}=0$, where $\nabla^{\mu}_g$ is the covariant derivative with respect to the metric $g_{\mu\nu}$.

We will assume that both metrics are homogeneous, isotropic and flat:
\begin{align}
 ds_f^2 &= Y(\tau)^2[-X(\tau)^2d\tau^2+\delta_{ij}dx^i dx^j], \label{sq}\\
 ds_g^2 &= a(\tau)^2[-d\tau^2+\delta_{ij}dx^i dx^j], \label{sg}
\end{align}
where $\tau$ is the conformal time, $a(\tau)$ is the scale factor of the space-time metric, and $X(\tau)$ with $Y(\tau)$ describe the evolution of the metric $f_{\mu\nu}$. Taking the matter sector to be a perfect fluid with energy density $\rho_0$ and homogeneous pressure $p_0$, we find the following equations of motion (which are analogous to the FRW equations in GR):
\begin{align}
\mathcal{H}^2& = \frac{a^2}{3}\left[\frac{\rho_0 }{M_g^2} +m^2\left(\beta_0 + 3\beta_1N+3\beta_2N^2+\beta_3N^3   \right) \right],\label{EqFried}\\
\mathcal{H}'&=\frac{a^2}{2}\left[-\frac{p_0}{M_g^2}-\frac{\mathcal{H}^2}{a^2} +m^2\left(\beta_0+\beta_1N\left[2+X\right]+\beta_2N^2\left[1+2X\right] +\beta_3N^3X \right)   \right],\label{EqDerH}    \\
h^2& =\frac{a^2}{3}\left(\frac{X^2}{N}\right)\nu^2\left(\beta_1+3\beta_2N+3\beta_3N^2+\beta_4N^3\right),  \label{Eqh2}\\
h'&=\frac{a^2}{2}\left[ \frac{2}{a^2}h_x h-\frac{h^2}{a^2} +\left(\frac{X}{N}\right)\nu^2\left(\beta_1+\beta_2N[2+X]+\beta_3N^2[1+2X]+\beta_4N^3X\right) \right],\label{EqDerh}
\end{align}
where it is implicit that all variables depend only on $\tau$, all primes represent conformal time derivatives, and we have defined $\mathcal{H}=a'/a$, $h=Y'/Y$, $h_x=X'/X$, $\nu=m/M_*$, and $N=Y/a$. Note that the parameter $M_*$ is redundant, as we can rescale the metric $f_{\mu\nu}$ to make $M_*$ take any value we want and redefine $\beta$s such that the action remains invariant. For simplicity, from now on we will use $M_*=1$.

\paragraph{Classes of solutions}
 From the Bianchi constraint\footnote{\color{black} The Bianchi constraint is obtained from the vanishing of the covariant derivative of the Einstein tensor, i.e., $\nabla^\mu G_{\mu\nu}=0$. This enforces that the sum of the remaining terms in the field equation vanishes under a covariant derivative (NB: terms may not vanish individually). The algebraic expression for this vanishing is the Bianchi constraint. }, one has
\begin{equation}
\left(X\mathcal{H}-h\right)\left(\beta_1+2\beta_2N+\beta_3N^2\right)=0,
\end{equation}
which leads to two possible classes of solutions: a trivial GR+$\Lambda$ class, and a class with separate expanding and bouncing branches.

The first class satisfies {{$\left(\beta_1+2\beta_2N+\beta_3N^2\right)=0$}}. This case leads to a constant $N=\bar{N}$,
such that $\mathcal{H}=h$, and the Friedmann equation becomes:
\begin{equation}
\mathcal{H}^2 = \frac{a^2}{3}\left[\frac{\rho_0 }{M_g^2} +\Lambda \right]; \; \Lambda= m^2\left(\beta_0 + 3\beta_1\bar{N}+3\beta_2\bar{N}^2+\beta_3\bar{N}^3   \right),
\end{equation}
which corresponds to general relativity with a cosmological constant. In this case, interactions between the perturbations of the $g$ and $f$ metrics vanish.

The second class satisfies {{$\left(X\mathcal{H}-h\right)=0$}}. This constraint can be used to find the following consistency equation:
\begin{equation}\label{Density}
\tilde{\rho} \equiv \frac{\rho_*}{m^2}=\frac{\beta_1}{N}+3\beta_2-\beta_0+3N(\beta_3-\beta_1)+N^2(\beta_4-3\beta_2)-N^3\beta_3; \quad \rho_*=\rho_0/M_g^2, 
\end{equation}
which relates $N$ and the density $\rho_0$. For a standard equation of state $p_0=w\rho_0$ (with $w$ constant), according to eq.~(\ref{Density}), at late times ($\tilde{\rho} \ll 1$), $N$ will approach a constant value, and both metrics enter an accelerated de-Sitter phase. However, at early times ($\tilde{\rho} \gg 1$), two types of behaviours can be identified: one where $N\ll 1$ (and $\beta_1\not=0$) and another where $N\gg 1$. The branch characterized by $N\ll 1$ at early times will be called {\it expanding branch}, as in this case both metrics expand with time (note that \cite{PhysRevD.90.124014} refer to this as the {\it infinite branch}). While the branch characterized by $N\gg 1$ will be called the {\it bouncing branch}, as in this case $g_{\mu\nu}$ expands but $f_{\mu\nu}$ bounces. 

\subsubsection{Linear perturbations}
Here we present the equations corresponding to the second branch of solutions above (since the first branch effectively reduces to $\Lambda$CDM). Define the perturbed metrics:
\begin{align}
 ds_f^2 &= Y^2[-X^2(1+2\phi_1)d\tau^2+2B_{1,i}X dx^id\tau  +[(1-2\psi_1)\delta_{ij}+2E_{1,ij}]dx^i dx^j], \label{spertq}\\
 ds_g^2 &= a^2[-(1+2\phi_2)d\tau^2+2 B_{2,i}dx^id\tau  +[(1-2\psi_2)\delta_{ij}+2 E_{2,ij}]dx^i dx^j], \label{spertg}
\end{align}
where $ds_f^2$ and $ds_g^2$ are the line elements for the metrics $f_{\mu\nu}$ and $g_{\mu\nu}$ respectively. We read from here that we have four scalar perturbation fields for each metric: $\phi_1$, $B_1$, $E_1$, $\psi_1$ for $f_{\mu\nu}$ and $\phi_2$, $B_2$, $E_2$, $\psi_2$ for $g_{\mu\nu}$, plus one  further scalar $\chi$ that describes perturbations of the perfect fluid \citep{MUKHANOV1992203}. We can set two of these variables to zero as a gauge choice. Using the Noether Identities approach (see \cite{PhysRevD.89.024034} and \cite{1475-7516-2014-12-026} for details), one learns that a sensible choice of gauge is to set $\psi_1=\chi=0$. The equation of motions for the seven remaining fields are, in Fourier space: 
\begin{align}
&2  \mathcal{H} \left(3\psi_2'+ k^2 E_2'\right)+\left( (1+w)\rho_*(3\psi_2+k^2E_2)+ m^2  NZ (3\psi_2+k^2(E_2-E_1))\right)a^2\nonumber  \\
& +2 \left(\psi_2 k^2+\mathcal{H}(3\phi_2\mathcal{H}-k^2 B_2)\right)=0 ,\label{G00}\\
&2(X+1)\psi_2'+2\mathcal{H}(X+1)\phi_2-m^2ZN(XB_1-B_2)+(1+w)\rho_*(1+X)B_2=0 ,   \label{G0i}\\
& 2 (k^2E_2^{''}+3\psi_2^{''}) + 2\mathcal{H}(3\phi_2'+6\psi_2'+2 k^2 E_2') -2k^2B_2' + 3 Za^2 m^2 N (\phi_1+\phi_2)X\nonumber\\
& + a^2\left( -3(1+w)\rho_*(2\phi_2+w(3\psi_2+k^2 E_2))  + 2 N m^2 (-3\phi_2 Z+(3\psi_2+k^2(E_2-E_1)) \tilde{Z}) \right) \nonumber\\
&+2(9 \mathcal{H}^2- k^2)\phi_2+2k^2(\psi_2-2\mathcal{H} B_2)=0 , \label{Gii} \\
& E_2^{''}-B_2'+2\mathcal{H} E_2'+(E_2 - E_1)a^2 m^2 N \tilde{Z}-\phi_2-2 \mathcal{H} B_2+\psi_2=0 \label{Gij}, \\
& 2 N  h k^2 E_1' -a^2 \nu^2 Z (k^2 E_2-k^2 E_1+3\psi_2) X^2-2 N  h k^2 B_1 X +6 \phi_1 h^2 N=0, \label{F00} \\
& 2h\phi_1N(X+1)+\nu^2 X a^2 Z(XB_1-B_2)=0,   \label{F0i} \\
& N X E_1^{''}- N (-2 X h +X')E_1'- X^2 \left( B_1' N + N\phi_1 X + 2NB_1 h+ \nu^2 a^2 \tilde{Z}(E_2-E_1)\right)=0,   \label{Fij}
\end{align}
where we have defined $Z=\beta_1+2\beta_2N+\beta_3N^2$, $\tilde{Z}=\beta_1+\beta_2N(1+X)+\beta_3N^2X$. We have omitted the explicit dependence of variables, but it should be clear that the perturbation fields depend on the conformal time $\tau$ and the wavenumber $k$.

From eqs.(\ref{G00})-(\ref{Fij}) we can see that the variables $B_1$, $B_2$, $\phi_1$ and $\phi_2$ never appear differentiated. This means that they are \textit{auxiliary variables}, and hence can be eliminated in favour of $E_1$, $E_2$ and $\psi_2$. Carrying out this exercise, one finds that $\psi_2$ also becomes an auxiliary variable. As a result, one reduces the system of equations down to just two true dynamical variables, $E_1$ and $E_2$. These both obey equations of the form:
\begin{equation}\label{EqE1E2}
E_a^{''}+c_{ab}E_b'+d_{ab}E_b=0,
\end{equation}
where $\{a,b\}$ can take the values $\{1,2\}$. The coefficients $c_{ab}$ and $d_{ab}$ depend on $k$, $\cal H$, $a$ and $N=Y/a$, and can be found in the appendix of \cite{1475-7516-2014-12-026}.

A similar analysis can be performed for vector and tensor perturbations about the homogeneous metrics of eqs.(\ref{sq}) and (\ref{sg}). The result is that the vector perturbations propagate three degrees of freedom, and the tensor sector propagates two d.o.f. Combined with the two scalar d.o.f.s this leads to a total of 2+3+2 =7 which, as explained above, is the correct number for a bigravity model.

We chose the above description as an illustration but we note that cosmological perturbations in bigravity models have been discussed prior to this in for example \cite{Comelli2012}.

\paragraph{Sub-horizon limit} 
Recall that the class of background solutions we have chosen to investigate here has two branches: one in which the $f$-metric expands along with the $g$-metric, and one in which it undergoes a bounce. The authors of \cite{1475-7516-2014-12-026} explored the solutions of eq.(\ref{EqE1E2}) and its vector and tensor equivalents, on both the expanding and bouncing branches, at early and late times, and on both super- and sub-horizon scales. We summarise their principal findings here.

The key result of their analysis was that on the expanding branch of solutions, the scalar perturbation $E_1$ suffers exponential growth on sub-horizon scales in both the radiation-dominated and matter-dominated eras. This represents a serious viability issue for the expanding branch of solutions, as exponential growth would rapidly cause metric perturbations to exit the linear regime and potentially impact structure formation. Furthermore, since the growth is exponential (rather than, say, power-law growth), it requires an extreme degree of fine-tuning of initial conditions to avoid this outcome. It is just possible that terms from higher-order perturbations conspire to prevent the exponential growth, but no concrete examples of this have been found. 

One may then look to the bouncing branch for a more viable cosmological solution. Indeed, the bouncing branch does have a viable scalar sector of perturbations, but \textit{only} under the condition $\beta_2=\beta_3=0$ in the bigravity action (eq.\ref{MGaction}); all other cases suffer an early-time subhorizon instability, similar to the expanding branch. Furthermore, although the linear scalar sector is free of instabilities under  $\beta_2=\beta_3=0$, it does not satisfy the Higuchi bound\footnote{The condition to have positive kinetic terms only in the action, see Appendix B of \cite{1475-7516-2014-12-026}.} that guarantees the ghost-freeness of the theory beyond the classical, linear regime. Hence the theory could still propagate ghost modes at perturbative orders beyond linear.

Under the restriction $\beta_2=\beta_3=0$, the vector and tensor sectors of the theory are found to possess power-law growing modes for certain parts of the remaining parameter space. These are not as damaging for the viability of the theory as exponentially growing modes, but further restrictions must be placed upon initial conditions and parameters of the model to ensure these growing modes do not harm the viability of the theory \citep{1475-7516-2015-05-052}.

\subsubsection{Parameters and prior ranges}

The parameter space to be constrained is that spanned by the $\beta_i$, where $i$ ranges from 0 to 4. In particular, the parameter $\beta_0$ acts like a cosmological constant for the regular metric, whilst $\beta_4$ can be interpreted as a cosmological constant for the $f$-metric. In this regard, then, there is a particular interest in finding viable models with $\beta_0=0$, as potential technically natural solutions to the cosmological constant problem. At the same time, models with $\beta_0\neq0$ and very small or zero values for the other parameters should be expected to provide a reasonable fit to data.

As we have seen above, on the bouncing branch the absence of exponential instabilities in the scalar sector enforces $\beta_2=\beta_3=0$. The remaining parameter space of $\{\beta_0,\beta_1,\beta_4\}$ has been initially explored by \cite{PhysRevD.90.124014}, who find two classes of viable models, one with $\beta_0\neq0$ and one with $\beta_0=0$. The latter of these they dub  `Infinite Branch Bigravity' (recall that their `infinite' branch is equivalent to our `bouncing' branch here). {\color{black} Since $\beta_0$ is effectively a cosmological constant for the g-metric, the authors focus on the infinite branch ($\beta_0=0$) in order to study a self-accelerating gravity model.} They place some simple constraints on this model using supernovae and growth rate data from the 6dFGS, BOSS, WiggleZ and VIPERS surveys.

They employ the quasi-static limit (with careful attention paid to the stability of the perturbations) to reduce the theory to a modified gravitational strength (which they denote $Y$) and a slip parameter ($\eta$). For early times these tend to $Y = 4/3$ and $\eta=1/2$, and hence they do not reduce to the standard $\Lambda$CDM values. The authors combined constraints on the $\beta_i$ parameters are presented as $\beta_1=0.48^{+0.05}_{-0.16}$ and $\beta_4=0.94^{+0.11}_{-0.51}$, with a best-fit value of $\Omega_{M0} = 0.16^{+0.02}_{-0.03}$ and an equation of state $w(z) \simeq- 0.79+0.21z/(1+z)$.

\subsection{Non-local gravity theories}
\label{sec:nonlocal}

Another way of modifying GR is to allow for the modified theory to have non-local terms. 
This non-locality was introduced using  different arguments based on quantum field theory grounds as in, for example, \cite{Deser2007NC,Deser2019NL} and \cite{Maggiore2014NL,Dirian2014CP,Belgacem2017NL,Belgacem2018II}. In the latter, it was argued that while the action of gravity is local, the corresponding quantum effective action, that includes the effect of quantum fluctuations, is  non-local. It is acknowledged there that these non-localities are understood at ultra-violet scales but not at the infra-red ones where they can have some contributions to gravity at cosmological scales. 

Such non-locality can be built phenomenologically using terms in the action of the form $\Box^{-1}R$ where $\Box$ is the D'Alembertian. In Fourier space, $\Box \sim k^2\sim 1/\lambda^2$, where $\lambda$ is the wavelength; this means these terms are significant on large scales with accompanying cosmological effects. To have some general insight into what $\Box^{-1}$ means, we need to think of it as the inverse operator of $\Box$, i.e., the Green function for $\Box$. This means that 
\begin{eqnarray}
{\Box^{-1}R}\sim \int d^4 x' G(x,x')R(x').
\end{eqnarray}
The operation requires an integral over some volume of spacetime and is thus non-local. The question of the boundary  of integration in such models is discussed in, e.g., \cite{Belgacem2017NL,Belgacem2018II}. 

\paragraph{Motivation} 
{\color{black} While this type of MG models received less attention when they were introduced some time ago, they have gained more popularity in the last decade because some of the recent models can exhibit cosmic acceleration without a cosmological constant, fit well cosmological observations, and are, in principle, indistinguishable from GR at small scales, i.e., galactic or planetary. The idea that non-local gravity might be a possible alternative to dark energy/cosmological constant was proposed by  \cite{Deser2007NC}; see further discussions in \cite{Woodard2014}. It was suggested there that, not only would it play a role on large scales, it might also explain the ``timing'' problem since the non-local corrections only come into play after radiation-matter equality. It turns out that the time dependence of the correction part is logarithmic in the matter era and leads to accelerated expansion today. 
The Deser-Woodard model contains a function $f(\Box^{-1}R)$ that can be mathematically tuned to mimic the background evolution of LCDM while, in principle, vanishing on small scales so the theory reduces to GR.
The more recent models by Maggiore and collaborators, see e.g., \cite{Maggiore2014NL,Dirian2015NG,Belgacem2017NL} are different from those of Deser and Woodard as they  involve a mass scale parameter associated with the non-local terms. These models make some predictions that make them close to the LCDM model but still  distinguishable from it. The mass scale is associated with a late-time effective dark energy density driving cosmic acceleration. The authors recognize the difficulty to derive the form of the non-local terms from a fundamental quantum field theory approach. Therefore they use a phenomenological approach strategy by investigating what desirable effects such terms must have on cosmological scales. Once such terms are found, one would go back to the more difficult problem of trying to derive this non-local term from fundamental grounds. Additionally, it was argued in \citep{Belgacem2017NL} that some of their models could have some specific predictions for the neutrino sector and reduce significantly some recent tensions between local and Planck measurements of the Hubble constant. Some extended discussions of non-local gravity that exhibit cosmic acceleration can be found in, e.g., \cite{Woodard2014,Belgacem2017NL,Koivisto2008NL}.
 
\subsubsection{Observational effects and status}
It turned out to be a non-trivial task to find non-local term that will: (1) behave like an effective dynamical dark energy producing an accelerated expansion without a cosmological constant, (2) have a well-behaved stable cosmological perturbation theory, and (3) fit well cosmological and astrophysical observations at the background level and the perturbation level in a competitive way compared to the LCDM model. 

The model from the proposal by \cite{Deser2007NC} that can mimic the LCDM background expansion is found to exhibit a weaker linear growth rate of large scale structure (LSS) and a stronger lensing power compared to the LCDM; see for example \cite{Amendola2019NL}. As for comparisions to the data, after some debate in \cite{Park2013SF,Nersisyan2017SF,Park2017RO}, it was  confirmed that the model is in agreement with LSS data. Most recently, \cite{Amendola2019NL} compared extensively the model to supernova, Planck CMB and redshift space distortion (RSD). They found that the model fits Planck CMB spectra and lensing data with a value of $\sigma_8$ that is significantly smaller than the one for the LCDM and similarly for the values of $f\sigma_8$ when redshift space distortion (RSD) data is added. It was found though that the model selection analysis only weakly favors LCDM over the non-local model. The authors highlight the importance of future lensing and galaxy clustering data in discriminating between the two models. 
 
Next, \cite{Maggiore2014NL,Belgacem2017NL,Dirian2015NG,MaggioreRT2014} proposed a few models that exhibit an interesting phenomenology and fit current cosmological data. They refer to one of their models as the $RR$-model which is based on using twice the Ricci scalar in the non-local action, i.e., $m^2 R\frac{1}{\Box ^2} R$, and a second model referred to as the $RT$-model which is based on using the Ricci scalar in the non-local part of the field equations and extracting its transverse part, i.e., $(\gmn \iBox_g R )^{\rm T}$.

\cite{Nesseris2014NL} have studied the cosmological perturbations of the $RT$-model and showed that it can provide statistically comparable fits to the CMB, BAO, SNIa and growth rate data as when the LCDM model is used. 
A similar extensive study for the $RR$-model has been worked out by \cite{Dirian2015NG,Belgacem2017NL} including  comparison to CMB, SNa, and BAO. It was found that the model is consistent with much of the existing cosmological data set and provides competitive fits to the data compared to the LCDM. However, when the RSD growth factor data, $f\sigma_8$, is added to the data sets, they find that the LCDM model fits the data marginally better.

While non-local models seem to have some success at the level of the cosmological data sets, it was very recently found that the $RR$ as well as the original Deser-Woodward models do not pass tests from Lunar Laser Ranging (LLR) that constrain tightly time variation in Newton constant, so they are essentially ruled out \citep{Belgacem2018LLR}. It remains open whether the Deser and Woodward model could have some physical screening mechanism to circumvent such a constraint, see comments in \cite{Amendola2019NL}.

However, \cite{Belgacem2018LLR} find that the $RT$ non-local model is not ruled out by the LLR bound. Also,  \cite{Deser2019NL} proposed an improved version of their non-local model with no-clear discussion yet of whether they pass the lunar ranging bounds.

\paragraph{Gravitational waves.}Finally, it was shown in \cite{Belgacem2017NL,Belgacem2018GW,Belgacem2018LLR,Amendola2019NL} that gravitational waves in the $RR$, $RT$, and Deser-Woodard models travel at the speed of light so the models pass the test of the binary neutron star merger GW170817 GW signal and its corresponding GRB 170817A electromagnetic signal.}

\subsubsection{Field and background equations}
{\color{black}
The field equations for the $RT$ model are given by 
\be\label{modelRT}
\Gmn -(1/3)m^2\(\gmn\iBox_g R\)^{\rm T}=8\pi G\,\Tmn\,,
\ee
where $m$ is a mass parameter and the factor $1/3$ is a normalization convenient in the 3-dimensional space. 

One can rewrite the field equations by defining an auxiliary field $U$ given by  
\be
U = -\Box^{-1}  R\, ,\label{Udef}
\ee
and a symmetric tensor 
\be
\Smn=-U\gmn =\gmn\ \iBox R,
\ee
which can be decomposed as
\be\label{defStran}
S_{\mu\nu}=
S_{\mu\nu}^{\rm T}+\frac{1}{2}(\n_{\mu}S_{\nu}+\n_{\nu}S_{\mu})\, 
\ee
with
$\n^{\mu}S_{\mu\nu}^{\rm T}=0$. 

The non-local field equation (\ref{modelRT}) can now be rewritten as \citep{MaggioreRT2014}
\bees
G_{\mu}^{\nu} + \frac{m^2}{3}  \[  U \delta_{\mu}^{\nu} + \frac{1}{2} 
\( \nabla_\mu S^\nu + \nabla^\nu S_\mu  \) \] &=& 8 \pi G T^{\mu}_{\nu}, \label{ModEE}\\
- \square_g U &=& R, \label{Loca} \\
 \nabla_\nu \( \nabla_\mu S^\nu + \nabla^\nu S_\mu  \) 
 &=& -2\partial_\mu U\label{DEEMcons}\, .
\ees
where \eq{DEEMcons} has been obtained by taking the divergence of \eq{defStran}. 

\cite{MaggioreRT2014} discussed the background evolution of this model in some detail. He used $x\equiv \ln a(t)$, a prime to denote $df/dx$ and $h=H/H_0$. He introduced the variable 
\be\label{defY}
Y=U-\dot{S}_0\,  
\ee
where ${S}_0$ is the t-component of $S_{\mu}$. The spatial 3-vector $S
^i$ vanishes since there is no preferred spatial direction in an FLRW model. 

The field equations give then the Friedmann equation
\be
h^2(x)=\Omega_M e^{-3x}+\Omega_R e^{-4x}+\g Y(x)
\, ,\label{hLCDM}\\
\ee
where 
\be
\g\equiv  m^2/(9H_0^2)\,,  
\ee
and an effective DE density can be defined as 
\be
\rde(t)\equiv \rho_0\g Y(x)\, , 
\ee
with $\rho_0=3H_0^2/(8\pi G)$. 

The evolution of the variable $Y(x)$ is given by the coupled system
\bees
&&\hspace*{-5mm}Y''+(3-\zeta)Y'-3(1+\zeta)Y=3U'-3(1+\zeta)U\, ,\label{sy1}\\
&&\hspace*{-5mm}U''+(3+\zeta)U'=6(2+\zeta)\label{sy3}\, ,\\
&&\hspace*{-5mm}\zeta(x)\equiv\frac{h'}{h}=-\, \,
\frac{3\Omega_M e^{-3x}+4\Omega_R e^{-4x}
-\g Y' }{2(\Omega_M e^{-3x}+\Omega_R e^{-4x}+\g Y)}\label{syz}\, .
\ees

Next, one can define a DE equation of state from the continuity equation 
\be\label{defwDE}
\dot{\rho}_{\rm DE}+3(1+w_{\rm DE})H\rho_{\rm DE}=0\, .
\ee
which can be written as 
\be\label{wg2}
w_{\rm DE}(x)=-1 -\frac{Y'(x)}{3Y(x)}\,  
\ee
by use of $\dot{\rho}=H\rho'$.

\subsubsection{Linear perturbations}
\cite{Nesseris2014NL} and \cite{Dirian2014CP} worked out linear perturbations for the $RT$ model in the Newtonian gauge with line element 
\be\label{defPhiPsi}
ds^2 =  -(1+2 \Psi) dt^2 + a^2(t) (1 + 2 \Phi) \delta_{ij} dx^i dx^j\,.
\ee

For the scalar sector, they expanded the auxiliary fields as
\be
U=\bar{U}+\d U\, ,\qquad S_{\mu}=\bar{S}_{\mu} +\d S_{\mu}\, .
\ee
In the isotropic FLRW background, the value $\bar{S}_{i}$ vanishes but the perturbation $\d S_i$ is a dynamical variable. Thus the scalar  perturbations  in this model are given by $\Psi,\Phi,\d U,\d S_0$ and $\d S$. 
The perturbed field equations are given in the appendix as well as their effective fluid representation.  

\paragraph{Sub-horizon limit} 

For Fourier modes well inside the horizon and relevant to LSS observations (i.e the large-$\hat{k}$ limit in \eqst{AppU}{AppZ}) from the appendix, one gets $\d U=2\Psi+4\Phi$ while $\d V={\cal O}(1/\hat{k}^2)$ and
$\d Z={\cal O}(1/\hat{k}^2)$. 

Therefore, from \eq{EE1xRT} one gets for sub-horizon modes \citep{Dirian2014CP},
\be\label{GeffGRTlargek}
\frac{G_{\rm eff}}{G}=1+{\cal O}\(\frac{1}{\hat{k}^2}\)\,. 
\ee
This result is consistent with the findings of \cite{Nesseris2014NL}. 

\citep{Dirian2014CP} showed using full numerical integration that $G_{eff}(z; k)/G$ for large $\hat{k}$ is equal
to one with great accuracy making the model compatible with galactic and solar tests such as the Lunar Laser  Ranging, unlike the $RR$ model.

\subsubsection{Parameters and prior ranges}

The viable $RT$ non-local model contains the same parameters as the standard flat $\Lambda$CDM model except for the derived parameter, $\Omega_{\Lambda}$, which is replaced by another derived parameter that is the mass parameter $m$. This parameter appears in the field equations and is embedded into the $\gamma$ parameter appearing in the modified Friedmann equation. It plays the role of an effective dark energy responsible for cosmic acceleration. So, comparing these models to observations requires changes in the background and growth equations, but does not require additional parameters compared to the standard model. 
It was found in \cite{MaggioreRT2014} that the $RT$  model effective dynamical dark energy can be matched to the observed value $\ode\simeq 0.68$ by tuning the mass $m$ by setting $m\simeq 0.67 H_0$. Once $m$ is fixed, the model predicts a dark energy equation of state of a phantom type with $w_0\simeq-1.04$ and $w_a\simeq -0.02$ \citep{MaggioreRT2014}, which is consistent with the Planck data.

Cosmological studies of this $RT$ model can be found in \cite{Nesseris2014NL} where the authors 
 compared them to CMB, BAO, SNIa and growth rate data. They found them statistically comparable to the LCDM model.}

\subsection{General $f(R)$ theories}
\label{sec:fr}

Models that involve higher powers of the Ricci scalar have been around almost as long as General Relativity \citep{Weyl:1918ib}. The most popular classification of these are known as $f(R)$-theories. They have been used as alternative ways of explaining the early-time acceleration already back in the seventies and eighties \citep{2012-Clifton-MG}, and recently as possible explanations for the late-time acceleration. The basic idea is to extend the Einstein-Hilbert action, which is linear in the Ricci scalar$(R)$, to depend on a more general function of $R$. This leads to fourth-order field equations. The generalization is obtained from a Lagrangian density of the form:
\begin{equation}
\mathcal{L} = \sqrt{-g} \,f\left(R\right)
\end{equation}

It is the simplest higher-order generalization to the Einstein-Hilbert density that is possible. For a more complete overview, see e.g., \cite{Felice2010FR,2012-Clifton-MG,2017-Burrage-Sakstein,Lombriser2014b,Capozziello2019}.

Although they are practically excluded by observational constraints, the reason to include $f(R)$ theories in our set of prioritised models is that they are so well-tested, and a wide range of both linear codes and N-body simulations can already produce mock data for these types of models. In general, we include $f(R)$ models mainly to ensure that we have a simple and testable beyond-$\Lambda$CDM model that we can easily compare results and constraints from LSST to.

\subsubsection{Observational effects and status}

Several different observational probes have been used to constrain $f(R)$ models. Primordial nucleosynthesis has been used to constrain the value of $f_R$ in the radiation epoch, and hence limits the number of starting points for the evolution of $f_R$ throughout later periods. Other probes that constrain $f(R)$ are looking at the growth of density perturbations in a flat universe. Both the matter power spectrum of the CMB and LSS can be used to put constraints of $f(R)$, \cite{2012-Clifton-MG}. Solar system probes and Dwarf galaxies have also been used for probing $f(R)$ and have found very tight constraints. $f(R)$ are  essentially ruled out observationally, see \cite{2017-Burrage-Sakstein,Lombriser2014b}. 

\subsubsection{Action and background equations}
There is in fact more than one way to approach describing the action for $f(R)$ theories \citep{2012-Clifton-MG}. One can take the metric variational approach, use the Palatini procedure, or the metric-affine approach. The last leads to an action of the form of:
\begin{equation}\label{fofRgen}
S = \int d\Omega \sqrt{-g}\left( R+ f(R)\right)+S_m\left(g_{\mu \nu}, \Gamma^{\mu}_{\nu \sigma},\Psi \right)
\end{equation}
where $S_m$ is the matter action of GR. Meanwhile the zeroth order Friedmann equations for the conformal Newtonian gauge can be found to be:
\begin{align}
H^2 &= \frac{1}{3f_R}\left[8\pi \rho - \frac{1}{2}\left(f-Rf_R\right)-3H\dot{f_R}\right] -\frac{\kappa}{a^2}\\
\dot{H} &= -\frac{1}{2f_R}\left(8\pi \rho + 8\pi P + \ddot{f_R}-H\dot{f_R}\right)+\frac{\kappa}{a^2}.
\end{align}
Alternatively, one can cast $f(R)$ as a special case of Horndeski, and its $\alpha$ parametrization. We will start with the form of the action commonly used for such models as well as the corresponding background equations derived from it, before showing the connection to the $\alpha$ parametrization. In the Einstein frame, such an action takes the form \cite{Hu:2007nk}:
\begin{equation}\label{actionHu}
\mathcal{S} = \int d^{4}x \sqrt{-g} \left[\frac{R+f\left(R\right)}{2 \kappa^2}+\mathcal{L}_m\right],
\end{equation}
with $\mathcal{L}_m$ being the matter sector Lagrangian, $\kappa^2=M^{-2}_p=8\pi G$ and with the function $f\left(R\right)$ now redefined (not to be confused with the one in equation (\ref{fofRgen})). Variation of this action gives the system of the modified Einstein equations for this class \cite{Hu:2007nk}:
\begin{equation}
G_{\mu \nu}+f_{R}R_{\mu \nu}-\left(\frac{f\left(R\right)}{2}-\Box f_{R}\right)g_{\mu \nu}-\nabla_{\mu}\nabla_{\nu}f_{R}=\kappa^2T_{\mu \nu},
\end{equation}
where by $f_{R}$ we denote the derivative with respect to the Ricci scalar $R$, $f_{R}=\frac{d f\left(R\right)}{d R}$. Taking the $0^{th}$ order perturbations, in the quasi-static limit, leads us to the corresponding background equation, the modified Friedmann equation \citep{Hu:2007nk}:

\begin{equation}
H^2-f_{R}\left(H H^{\prime}+H^2\right)+\frac{1}{6}f\left(R\right)+H^2f_{RR}R^{\prime}=\frac{m^2 \rho}{3},
\end{equation}
where $R=12H^2+6HH^{\prime}$ and $^{\prime}$ is used to denote $\frac{d}{d\ln a}.$
It is worth noticing at this point that in the limit that $f\left(R\right)$ becomes a constant, the above equations reduce to the familiar set describing a $\Lambda$CDM evolution. Any well-defined $f\left(R\right)$ function contains in principle enough freedom to mimic the expansion history by any value of $w$, as desired, which implies that studying the linearized perturbations is necessary to break the degeneracy with $\Lambda$CDM, which we will do in the following section.

\subsubsection{Linear perturbations}
Since linearized $f(R)$ theories fall within the Horndeski class, they can be treated using the formalism in Section \ref{sec:horndeski:implementation}. In terms of the $\alpha$ parametrization and as pointed out in \cite{Bellini14}, metric $f(R)$ theories correspond to $\alpha_K=\alpha_T=0$ and $-\alpha_B=\alpha_M=B \frac{\dot{H}}{H^2}$, where B is a useful parameter defined as:
 \begin{equation}
 \label{BfR}
B=\frac{f_{RR}}{1+f_{R}}R^{\prime}\frac{H}{H^{\prime}},
\end{equation}
with $f(R)$ defined as in equation (\ref{actionHu}). As shown previously, in the quasi-static limit and for sub-horizon scales, the linear perturbations in the Newtonian gauge are given by
\begin{align}
\label{frPoisson}
-2\frac{k^2}{a^2}\Psi&=8\pi G_{\rm eff} (z) \rho_m\delta_m =-2\frac{k^2}{a^2}\Psi_{GR}\times\frac{G_{\rm eff} (z)}{G_N}
\end{align}
and
 \begin{align}
 \label{frslip}
-\frac{k^2}{a^2}\left(\Phi+\Psi\right)  = -\frac{k^2}{a^2}\Psi_{GR}2\times \left[1+\Sigma(z)\right].
\end{align}
Under the simplifications introduced by the vanishing of $\alpha_K$ and $\alpha_T$, and the fact that $\alpha_M=-\alpha_B$ in this class, we get that:
\begin{align}
\frac{G^{k\rightarrow\infty}_{\rm eff}(z)}{G_N}&=\left(\frac{M_P^2}{M_*^2}\right)\left(1+\frac{ \alpha^2_M }{2 \DD\,  c_s^2}\right)
 \label{frG}
\end{align}
Using eq.(\ref{cvb}), this corresponds to $\mu(z)=\alpha^2_M /2 \DD\,  c_s^2$. From eq.(\ref{yui}) we also find:
\begin{align}
\Sigma^{k\rightarrow\infty}(z)&= 0
 \end{align}
 For these models, we have that:
 \begin{align}
\DD\,  c_s^2 = -\frac{\left(2+\alpha_M\right)\left[\dot{H}-H^2\frac{\alpha_M}{2}\right]+H\dot{\alpha}_M+\frac{\rho_m+ p_m}{M_*^2}}{H^2}.
 \end{align}

\subsubsection{Parameters and prior ranges}

Among the current models we have the following that attempt to avoid many of the short-comings given by the simpler models.

\cite{Starobinsky:2007hu}:
\begin{equation}
f(R) = R-\mu R_c \left[1 - \left(1+\frac{R^2}{R_c^2}\right)^{-n}\right].
\end{equation}
\cite{Hu:2007nk}: This model is by far the most popular and best studied $f\left(R\right)$ model in the literature, that has been simulated by a variety of N-body codes \cite{Winther:2015wla}. It is characterized by a broken power-law function of the form
\begin{equation}
f(R) = -m^2\frac{c_1 \left(\frac{R}{m^2}\right)^n}{c_2 \left(\frac{R}{m^2}\right)^n+1},
\end{equation}
where $m^2=\frac{\kappa^2 \rho_m}{3}$ and $c_1$, $c_2$ and $n$ are free parameters. The number of free parameters can be further reduced since, in order to match the $\Lambda$CDM expansion history, we should have $\frac{c_1}{c_2}=6\frac{\Omega_{\Lambda}}{\Omega_{m}}$. In addition, evaluated at $z=0$, the scalaron $f_R$ becomes
\begin{equation}
f_{R0} = -\frac{c_1}{c^2_2}\left(\frac{12}{\Omega_{m}}-9\right)^{-n-1},
\end{equation}
which means that the model can be fully characterized by specifying the values of $n$ and $|f_{R0}|$, which is the most common parametrization of this subclass. The linearized perturbations in the Newtonian gauge give the modified system of Poisson and Klein-Gordon equations \cite{Zhao2011Z}:
\begin{equation}
-\frac{k^2}{a^2}\Psi=4 \pi G \delta \rho_m + \frac{1}{2}\frac{k^2}{a^2}\delta f_{R}
\end{equation}
and
\begin{equation}
\left(k^2+a^2\mu^2\right)\delta f_{R}=\frac{8 \pi G}{3}a^2  \delta \rho_m,
\end{equation}
with $\mu$ being a characteristic mass given by
\begin{equation}
\mu=\left(\frac{-m^2\left(a^{-3}+4\frac{\Omega_{\Lambda}}{\Omega_{m}}\right)}{(n+1)|f_{R0}|}\left(\frac{a^{-3}+4\frac{\Omega_{\Lambda}}{\Omega_{m}}}{1+4\frac{\Omega_{\Lambda}}{\Omega_{m}}}\right)\right)^{\frac{1}{2}}.
\end{equation}
The above system of equations is, as expected, a special case of equation (\ref{frPoisson}) that is general for all $f(R)$ models.
\newline
\cite{Appleby:2007vb}:
\begin{equation}
f(R)= R_c \log \left[e^{-\mu} + \left(1-e^{-\mu}\right)e^{-R/R_c}\right].
\end{equation}

\subsection{N-body simulations for Beyond $w(z)$CDM models}
\label{sec:nlsims}
There is much knowledge to be learned from studying the non-linear evolution of the universe. Deviations from GR on linear scales are now quite tightly constrained, forcing a viable appearance of new gravitational physics to take place predominantly in the non-linear regime.

Given the difficulty of solving the relevant equations, non-linear evolution is best examined via N-body simulations. However, there is a series of challenges connected to simulating `Beyond $w(z)$CDM' theories in N-body simulations. At present we lack a unique way to connect linearized parameterizations of MG theories to a non-linear counterpart parameterization. Constructing parameterizations for non-linear scales has proved difficult, and no established solution exists at the time of writing. Even if such a formalism is developed in future, the mapping between linear and non-linear parameters is unlikely to be one-to-one. That is, theories with different non-linear behaviours could match onto the same set of linear parameters. Because of this issue,  it will be challenging to make consistent MG linear + quasi-linear/non-linear predictions like we can for GR+$\Lambda$CDM. 

Other challenges are the non-locality or non-linearity of some modified gravity models, which prevent approximations normally used in N-body simulations from being feasible. Furthermore, evolution of the extra fields present in nearly all MG theories means that more processing time per timestep is needed, and results in an increased memory load. 

We choose to strike a balance between investing in new simulations that match our prioritised MG models, and using existing expertise and publicly-available simulation codes. Some existing codes, though they may simulate lower-priority models (according to the criteria laid out earlier in this document), have interesting behaviors on non-linear scales, e.g., screening. By looking for generic features of screening we hope to use them to help develop possible model-independent tests.

\subsubsection{Currently available simulations}
The latest overview of modified gravity models implemented in N-body simulations comes from the modified gravity code comparison project \citep{Winther:2015wla} (see references therein for details of individual codes). This shows us that most codes so far have implemented variations of $f(R)$, symmetron and DGP models, which are representative examples of the chameleon (and phenomenologically similar symmetron) and Vainshtein screening mechanisms. The non-linearities these theories introduce to the Klein-Gordon equation guarantee their phenomenological viability in the high-density environments, but render their N-body simulations more computationally expensive compared to their $\Lambda$CDM counterparts. As a result, besides the exact simulations mentioned above, several efficient methods have been considered in the literature, designed for use in cases where computational resources are limited.

Such an efficient method is the COLA (COmoving Lagrangian Acceleration) hybrid scheme, developed in \cite{Tassev:2013pn} for $\Lambda$CDM, which evolves the linear scales analytically using $2^{nd}$ order Lagrangian perturbation theory and employs a full N-body solver only for the smaller scales. The COLA method was extended to incorporate chameleon and symmetron screening mechanisms in \cite{Valogiannis:2016ane}, combined with a phenomenological screening implementation, and was shown to produce accurate and fast results for a wide range of k modes. \cite{Winther:2017jof} showed that it also works for the Vainshtein mechanism and in particular the nDGP model. The code used by \cite{Winther:2017jof} is available to the public. 

\subsubsection{Going forward}
A simulation of our top-priority model, Horndeski gravity (section \ref{sec:horndeski}), is highly desirable for tests of gravity with LSST. Discussions regarding how to best tackle this considerable task are ongoing within our Beyond $w(z)$CDM taskforce. At present, our plan is to start by implementing a framework based on the $\mu$ and $\Sigma$ parametrization (presented in \ref{sec:musigma}). The ultimate goal is to upgrade this to an implemention of the $\alpha_i$ parametrization as described in \ref{sec:horndeski:implementation}. That said, even an implementation of the quasistatic $\mu$ and $\Sigma$ parameters would serve as a useful new tool for our work. 
{\color{black}We would of course have to consider how to deal with screening, most likely we would have to come up with a way to manually turn on a screening effect at a certain distance.}

Ideally, we would like to perform enough N-body simulations that will allow the creation of an emulator, that effectively interprets over the  parameter space (be it $\mu$ and $\Sigma$ or the $\alpha_i$ parameters).
Such a task will most likely be very computationally expensive. An alternative is to build a more efficient simulation that applies the COLA method (see above) to Horndeski gravity, and verify the theoretical accuracy of this using a reduced number of representative `full' Horndeski simulations. We could then construct an emulator from the COLA-Horndeski simulations, resulting in viably efficient software with which to generate predictions, and lower computational costs. In addition we will use the full N-body simulations of Horndeski gravity, to create mock-galaxy catalogs that can be used by the entire collaboration to prepare for the analysis of LSST data.

Well-established simulations of $f(R)$ gravity, one of our second tier prioritised models, already exists. We will use these to perform a small number of model-specific tests, targeting the most up-to-date parameter ranges of interest. This will yield a comparison of the errors achievable in parameterized versus model-specific tests.

\section{Phenomenological parameterizations}
\label{sec:parameterizations}
Although the main objective of this document is to provide a set of recommendations on specific MG models or classes of MG models, we provide here a brief discussion of  phenomenological parameterization methods. In such approaches, one considers instead general parameterizations which encompass the behavior of multiple MG theories. We discuss two common approaches of this type. 

\subsection{$\mu$ and $\Sigma$ parameterization}
\label{sec:musigma}
The functions $\mu$ and $\Sigma$ have already been introduced in Section \ref{sec:horndeski:implementation} in the context of the quasistatic limit of Horndeski theories. Alternatively, $\mu$ and $\Sigma$ can be thought of as a means to parameterize phenomenological modifications to gravity, wherein $\mu$ represents changes from the General Relativistic clustering of matter, and $\Sigma$, deviations from the standard behavior of gravitational lensing (see, e.g., \citealt{DESY1ext, Joudaki2017, Planck2015MG}). In this section we review that approach.

With the scalar perturbations in Newtonian gauge [eq.\,\eqref{lineel}], in the sub-horizon regime the relation between the gravitational potential $\Phi$ (the spatial component) and the matter overdensity in GR is given by a Poisson equation \cite{Ma1995ApJ}
\begin{equation}
-\frac{k^2}{a^2}\Phi=4\pi G_N\rho_m\delta_m\label{eq-Poisson-GR}\,.
\end{equation}
Note that in GR although the Poisson equation is sensitive only to $\Phi$, it is $\Psi$ (the temporal component) that determines the equation of motion of non-relativistic particles and hence the growth equation. In the absence of pressure anisotropy we have $\Psi=\Phi$, so the Poisson equation is often also expressed with $\Psi$ especially at late times. The combination $\Psi+\Phi$ ($\equiv2\Psi_{\rm w}$, where $\Psi_{\rm w}$ is called the Weyl potential) affects the motion of relativistic particles. Many modified gravity theories, such as the Horndeski family discussed in Sec.\,\ref{sec:horndeski}, modify the motion of non-relativistic and relativistic particles respectively in broadly similar ways. In the quasi-static limit, such MG effects can be expressed as two modified Poisson equations,
\begin{align}
-\frac{k^2}{a^2}\Psi&=4\pi (1+\mu)G_N\rho_m\delta_m\label{eq-Poisson-Psi-MG}\,,\\
-\frac{k^2}{a^2}(\Psi+\Phi)&=8\pi (1+\Sigma) G_N\rho_m\delta_m\label{eq-Poisson-Weyl-MG}\,.
\end{align}
Note that, compared to eqs.(\ref{bnm}) and (\ref{rty}), here we have taken $M_*=M_P$ for simplicity. Due to the effects of $\Psi$ and the Weyl potential $\Psi_{\rm w}$ on non-relativistic and relativistic particles respectively, we can constrain $\mu$ predominantly by the rate of linear structure growth and $\Sigma$ by gravitational lensing.\footnote{Gravitational lensing is affected to a lesser extent by $\mu$ as well, via a modified matter power spectrum; see Eq.\,\eqref{eq-WL-other-cross} later in this section.} We follow the convention of \cite{DESY1ext} and allow both $\mu$ and $\Sigma$ to vanish in GR. 

In general, $\mu$ and $\Sigma$ can be both time- and scale-dependent. It has been shown that the scale dependence is sub-dominant, at least for near-future surveys \citep{Baker14}, therefore it is common to choose for them a scale-independent ansatz when performing analysis. Choosing a fiducial time-dependence for $\mu$ and $\Sigma$ is less straightforward. As the parameterization is intended as a general phenomenological case to encompass multiple theories, there is no clear choice of functional form. It is common in current works (see, e.g., \cite{ZhaoEtAl2010,2013MNRAS.429.2249S} and many others), to model $\mu$ and $\Sigma$ as proportional to the fractional energy density of the effective dark energy component, an ansatz which derives from the heuristc argument that we expect modifications to gravity in general to become non-neligible at late times. However, this is just one possible choice, and an active exploration of more complex or better-suited choices is ongoing in the literature; we refer the reader to some of the reviews or comparisons of such parameterizations, see e.g., \cite{Daniel2010MG,Ishak2018}.

Because of the first modified Poisson equation \eqref{eq-Poisson-Psi-MG}, the equation for the late-time linear structure growth is changed in this parameterization from its GR counterpart, to read
\be 
\label{eq-MG-growth}
\frac{d}{da}\left(a^3H(a)\frac{dD}{da}\right)=\frac{3(1+\mu)}{2}\Omega_m(a)aH(a)D\,,
\ee
where $D$ is the growth factor. The above equation reduces to the GR case when $\mu=0$. We call the solution of the above equation $D_{\rm MG}(a)$ and the GR solution $D_{\rm GR}(a)$. Because MG effects are assumed to become important only at late times, we impose an asymptotic boundary condition $D_{\rm MG}(a\rightarrow0)=D_{\rm GR}(a\rightarrow0)=a$. Note that, by enforcing $D_{\rm MG}(a)$ and  $D_{\rm GR}(a)$ to approach $a$ at early times, neither is normalized to unity today. 

In the simple case in which $\mu$ is scale-independent, we are in the linear and quasistatic regime, and MG effects become important only at late times, the MG matter power spectrum can be related to the GR matter power spectrum via 
\be
\label{eq-MG-GR-matter-power} 
P_m^{\rm MG}(k,a)=P_m^{\rm GR}(k,a)\times\left[\frac{D_{\rm MG}(a)}{D_{\rm GR}(a)}\right]^2\,.
\ee
Note that in particular, if the MG effects become important before the growth equation approximation\footnote{The growth equation approximation refers to the situation where the fractional density perturbations of cold dark matter and baryons become equal, and the evolution of the total matter overdensity is captured by the growth factor govern by Eq.\,\eqref{eq-MG-growth} (with $\mu=0$ in GR). In GR, this approximation becomes valid at times sufficiently later than last scattering, including the time range of LSS observations.} becomes valid, the above treatment would not be correct. 

Within the simplified regime described above, the other MG variable $\Sigma$ does not affect the linear matter growth, but it affects the magnitude of the Weyl potential. Therefore, it changes the geodesics of photons and hence lensing effects.  For example, within the Limber approximation, the angular power spectrum of the two-point cross-correlation between galaxy weak lensing ($L$) and any other probe ($a$, including $L$) can be written as
\be\label{eq-WL-other-cross}
C_{\ell}^{La}=\frac{2\ell}{2\ell+1}\int^{\chi_\infty}_0 \frac{d\chi}{\chi^2} P_m\left(\frac{\ell}{\chi},z(\chi)\right)\tilde{\Delta}_{\ell}^L(\chi)\tilde{\Delta}_{\ell}^a(\chi)\,,
\ee
where $P_m$ is the matter power spectrum and is $P_m^{\rm MG}$ if $\mu$ is non-zero. Due to the modified Poisson equation \eqref{eq-Poisson-Weyl-MG}, the function $\tilde{\Delta}_{\ell}^L(\chi)$ in MG reads,
\be\label{eq-Delta-L-GR}
\tilde{\Delta}_{\ell}^L(\chi) = \frac{3}{2}\Omega_{m}^0H_0^2\sqrt{\frac{(\ell+2)!}{(\ell-2)!}}\frac{(1+\Sigma)\chi}{\ell^2}(1+z(\chi))W^L(\chi)\,,
\ee
where $W^L$ is the lensing kernel.  Both cosmic shear and CMB lensing, for example, can be used to constrain $\Sigma$. 

{\color{black} The parameters of various functional ansatzes for $\mu$ and $\Sigma$ have been constrained with existing cosmological data from numerous surveys (see, for example, \cite{2013MNRAS.429.2249S, Joudaki2017, 2018arXiv180706209P}). For example, a recent set of constraints which used galaxy clustering and weak lensing measurements from the first year of the Dark Energy Survey in combination with complementary external data sets found that for $\mu$ and $\Sigma$ $\propto \Omega_{\Lambda}(z)$, ${\mu_0}=-0.11^{+0.42}_{-0.46}$, ${\Sigma_0}=0.06^{+0.08}_{-0.07}$ (\cite{DESY1ext}). We note that in order to effectively constrain $\mu$ with LSST, measurements of redshift-space distortion from contemporaneous spectrosopic galaxy surveys (e.g., the Dark Energy Spectroscopic Instrument, the 4 Meter Multi-Object Spectroscopic Telescope) will be helpful in breaking the degeneracy which arises between a scale-independent $\mu$ and the galaxy bias. }


\subsection{The $E_G$ statistic}
\label{sec:EG}
The use of the $\mu$ and $\Sigma$ parameterization in observational tests of gravity is predicated on the assumption that the analysis will consist of making constraints on parameters of these functions. Another approach is to strategically combine observables into a single statistic in such a way as to null dependence on particular sources of systematic errors. The most well-explored statistic of this type is called $E_G$, which is constructed to be independent of galaxy bias. 

Originally defined in \cite{Zhang2007}, $E_G$ combines galaxy-galaxy lensing and spectroscopic galaxy clustering measurements, including redshift space distortions, to probe gravity while limiting sensitivity to a linear galaxy bias and the cosmological parameter $\sigma_8$. Although it was first defined in Fourier space, a real-space version of the definition is more commonly used in practice \citep{Reyes2010}:
\begin{equation}
E_G(R) = \frac{\Upsilon_{gm}(R)}{\beta \Upsilon_{gg}(R)}
\label{eg}
\end{equation}
where $R$ is projected radial separation, $\beta$ is the linear growth rate of structure ($f$) divided by the linear galaxy bias, $b$, and $\Upsilon_{gm}(R)$, $\Upsilon_{gg}(R)$ are modified versions of the conventional observables of galaxy-galaxy lensing and galaxy clustering respectively, altered to limit sensitivity to effects below a certain projected radius. Because $\Upsilon_{gg} \propto b^2$, $\Upsilon_{gm} \propto b$ and $\beta = f / b$, on sufficiently large linear scales, where galaxy bias is linear, it cancels. It is important to note that this requires that the lens galaxies used in the galaxy-galaxy lensing measurement are the same population as used for the galaxy clustering measurement.

Lensing measurements are sensitive to the Weyl potential $\Psi_{\rm W} =\frac{1}{2}( \Psi + \Phi)$, while the clustering of overdensities is related to $\Psi$. It can thus be seen schematically that
\begin{equation}
E_G \propto \frac{\Psi + \Phi}{\Psi}.
\end{equation}
In GR, and at large linear scales, $E_G$ is predicted to take the scale-independent value of $\Omega_M / f(z)$, where $z$ is the effective redshift of the spectroscopic galaxy sample. Measuring $E_G(R)$ on large scales thus can act as a consistency check for GR,{\color{black} and numerous such measurements have been made using available spectroscopic samples at multiple redshifts \citep{Reyes2010, Blake2015, Alam2017, Amon2018, Singh2018}, all thus far showing consistency with GR predictions.} 

\section{Systematic effects and deviations from GR}
\label{sec:systematics}


With increasing statistical power of ongoing and upcoming surveys, cosmological constraints have become systematic-error limited. This will certainly be the case for LSST, which will statistically constrain MG parameters to the percent or sub-percent level (see for example \cite{Ferte2017}). {\color{black} LSST cosmic shear and galaxy-galaxy lensing measurements will be subject to numerous sources of systematics effects, including measurement systematics such as point-spread function (PSF) contamination, shear estimation uncertainties, and uncertainties in photometric redshifts (see, e.g., \citep{Weinberg2013,Mandelbaum2018}, as well as modeling systematics such as the intrinsic alignment (IA) of galaxies (see, e.g., \citep{Troxel15,Kirk15}), galaxy bias, baryonic effects (see, e.g., \citep{Chisari2018,Huang2018}) and the modeling of the matter power spectrum on small (nonlinear) scales. Incomplete or improper treatment of these systematic effects can bias the inferred cosmological parameters and, in some cases,  can potentially mimic the effect of new fundamental physics, including modifications to gravity or non-standard dark energy}. In the context of constraining or measuring new fundamental physics, this challenge is compounded by the fact that astrophysical systematics, including IA, baryonic effects, and the nonlinear matter power spectrum may be different than in a $w(z)$CDM universe.

Let us first discuss in detail the systematic effect of IA, as an example to guide the broader question of systematic effects in tests of beyond $w(z)$CDM cosmology. There are two types of IA that contaminate cosmic shear. The first is due to the intrinsic ellipticity auto-correlations that are present between close galaxies that formed in the same tidal gravitational field. These are called the II for the 2-point correlations and III for the 3-point functions. The second type is due to a correlation/anti-correlation between a galaxy tangentially lensed by a foreground structure and a galaxy near this lens structure oriented radially towards it \citep{Hirata04}. These are referred to as gravitational shear - intrinsic ellipticity or the GI for the 2-point correlations and GGI, GII for the analogous 3-point functions. While II-type correlations can be significantly reduced by removing or down-weighting nearby pairs of source galaxies (e.g.,  by considering only cross-bin pairs), the GI-type cannot be eliminated by such a technique as they are present at large line-of-sight separations as well.

Failing to account for IA has been shown to bias the accuracy of the dark energy equation of state parameters by up to $50\%$ (for an LSST-like survey) \cite{Bridle07,Ji17} when cosmic shear is used alone. IA mitigation methods such as marginalization over IA parameters \citep{Krause16,Blazek2017} or self-calibration techniques \citep{Zhang08,Troxel12,Ji17} have been shown to be able to remedy to the majority of such biases, but practical challenges remain in the context of MG \citep{Laszlo2012,Kirk13}. IA can affect the measured shear power spectra and the strength of the growth of large scale structure as inferred from these spectra. This change is potentially degenerate with the effects that deviation from GR can have on the growth and could mimic MG. Furthermore, intrinsic alignments themselves are a gravitational effect: galaxy shapes are determined by their local tidal fields and formation histories. To be fully consistent, we would therefore need to model intrinsic alignments themselves within MG. For instance, in the $\mu$ and $\Sigma$ parameterization (Sec.~\ref{sec:musigma}), $\mu$ alters the Poisson equation for non-relativistic particles and thus will impact IA, while $\Sigma$ alters the trajectories of relativistic particles and thus the lensing amplitude. Some initial work using earlier lensing data from CFHTLenS showed no significant correlations between the amplitude of IA and MG parameters \cite{Dossett15}, although a limitation to these results was that the IA model assumed $w(z)$CDM. Such a correlation remains to be studied using a full MG parameterization and for more precise surveys such as LSST.

Another significant source of systematic error which is particular to the MG scenario is uncertainty in modeling the non-linear growth of structure. In GR, we typically use fitting formulae calibrated from N-body simulations \citep{Smith2003, Takahashi2012} to make theoretical predictions of cosmological observables at the relatively small scales where non-linear physics is required. Constructing an equivalent formula for each modified gravity scenario of interest {\color{black} remains a challange} due to the computational investment required, and some of the more general parameterizations introduced above (e.g., the quasistatic parmeterisation) are only well-connected with viable theories of gravity at linear scales. A straightforward approach to reduce biases due to uncertainty in non-linear effects in MG is to enforce a scale cut based on the estimated fractional error in modeling, e.g.,  \cite{Leonard2015EG}. A more precise approach, typically used by current experiments, is to cut the scales that would bias parameter inference at a given level compared to the statistical uncertainty -- the scales to be cut then depend on the statistical power of the survey. With LSST, this strategy may require cutting an impractical amount of data from the analysis. Similar considerations apply to the treatment of baryonic effects. If these effects on small scales are ``separable'' from the impact of MG, it is possible that the same mitigation approaches used in $w(z)$CDM analyses will remain valid (see, e.g., \cite{Mead2016} for a halo model treatment of both baryonic and MG effects). 

 However, there is no guarantee of such separability. Complete simulations, including both modified gravity and baryonic effects, require substantial resources, likely prohibitive for DESC (see Sec.~\ref{sec:nlsims}) {\color{black}but worth exploration. Indeed, it is worth noting that for current and planned surveys over half of the signal-to-noise lies within the non-linear regime. So discarding such scales leads to significant loss in constraining power and degradation in parameter constraints. It is thus worth pursing some ongoing efforts and success in simulating Vainshtein or Chameleon mechanisms using efficient methods such as COLA approaches and emulators as we discuss in section \ref{sec:nlsims}}.

  Considerable work is still required to better understand nonlinear scales in MG and to develop methodologies for incorporating these scales in MG analyses in the presence of astrophysical systematics.

{\color{black}We did not discuss here systematics affecting redshift-space galaxy clustering since LSST is a photometric redshift  survey.} 

To conclude, we emphasize that careful forecasting must be done for LSST-DESC analyses in order to assess how these systematic effects will impact constraints on MG parameters. The results of such forecasts will provide further guidance on where additional effort is most critical and which regions of MG model space are most accessible to LSST.


\section{Additional information}
\label{sec:additional}

\subsection{Selection process, criteria used and ranking}
\label{sec:criteria}

Table \ref{Table-MG-criteria} outlines the criteria used in our model selection process. We briefly justify here why five models are selected as highest priority according to this grading. It should be stressed, though, that there is an unavoidable element of subjectivity to this process. As noted in the table caption, a lower score indicates a more favorable model.  

\paragraph{Horndeski: score ~2.} The Horndeski family, and by implication its extensions, were selected as our top priority for Beyond $w(z)$CDM investigations of LSST. This is due to their reasonable physical motivation (as the most general theory of a scalar and a metric with second-order equations of motion). The family of theories is stable under a few mild restrictions. Patches of viable parameter space remain, though are reduced in size following the binary merger event GW170817 (assuming no loopholes along the lines of \cite{BattyeGW2018,deRhamGW2018}). 
Thanks to the existence of the hi$\_$class code \citep{Zumalacarregui:2016pph}, many observables can already be calculated. There are currently no N-body simulations of Horndeski, but we see no fundamental reason why this cannot be done in future (particularly via an emulator approach). 

\paragraph{Bigravity: score ~4.} Well-motivated as a general, ghost-free theory of a tensor field interacting with the metric (somewhat akin in spirit to a spin-2 version of Horndeski above). However,  it has become apparent that many parts of the parameter space are plagued by linear instabilities, either in the scalar or tensor sectors (see section \ref{sec:bigravity}). It is in principle possible that these can be cured in a higher-order perturbative analysis, but at present the situation does not look promising. Background and linear calculations are available, but nonlinear calculations and simulations are not. 

\paragraph{Non-local gravity: score ~4.} Interesting developments with models fitting current data and self-accelerating. Some motivations from quantum spacetime ideas and emergence are interesting conceptually, but not concretely realised at present. Solid predictions are available and viable for background and linear observables, with the same number of parameters as $\Lambda$CDM. Predictions close to $\Lambda$CDM model but different enough and should be detectable by future surveys such as LSST and Euclid. The theory has luminal gravitational wave speed and passes the GW170817/GRB 170817A constraint. Codes for distances, large-scale structure and full CMB is publicaly available via Github repositories. We are unaware of any nonlinear calculations at this moment although we are aware that N-Body simulation work is ongoing within the group of Maggiore (private communication). 

\paragraph{$f(R)$ gravity: score ~4} $f(R)$ gravity has a sensible motivation as the next-order strong-field correction to GR, and is stable under a few mild restrictions on the form of $f(R)$. It can also naturally incorporate chameleon screening. However they are essentially ruled out by observational constraints. Calculations are mature and N-body simulation codes are available so these models are often used for pipeline and framework testing and we kept them here for that purpose for DESC studies. 

{ \color{black} \paragraph{parameterized Poisson-slip approaches: (no score).} Although strictly speaking a parameterization of models rather than a model itself, we include this methodology just for completeness and clarification. We do not attempt to score this approach according to the criteria of Table \ref{Table-MG-criteria}, since the parameterization describes many models with differing scores. Instead we note that this two-function framework serves as a valuable first step towards the more sophisticated Horndeski formalism, and hence is a natural choice for LSST implementation. The formalism maximises on simplicity and efficiency, at the expense of restricted applicability (the quasistatic regime only), and an obscured connection to the modified gravitational Lagrangian. }

\begin{table}[htbp!]
\scriptsize
\centering
\begin{tabular}{|c|>{\raggedright}p{0.16\textwidth}|>{\raggedright}p{0.16\textwidth}|>{\raggedright}p{0.16\textwidth}|>{\raggedright}p{0.16\textwidth}|>{\raggedright}p{0.16\textwidth}|}
\hline
\hline
Grading &
Physical motivation &
Viability\newline + Consistency & 
Maturity\newline + Calculability &
Information content      &
Accessibility to LSST   \tabularnewline
\hline
1& 
Well-motivated, including stability \& lack of ghosts &
Interesting model parameters not yet ruled out; all LSST observables can be calculated consistently &
Can calculate anything that can be calculated in LCDM+GR &
General features say something substantial about new fundamental physics &
Decisive constraints possible with LSST (possibly when combined with other surveys) \tabularnewline

\hline

2 & Interesting "why not" physics, new terms/interactions stable without miraculous cancellations or fine-tunings &
Interesting model parameters not yet ruled out; multiple LSST observables can be calculated consistently &
Full non-linear N-body calculations possible &
General features say something substantial about new physics &
Strong constraints from more than one LSST observable \tabularnewline

\hline

3&
Maybe not natural, but addressing a specific observational anomaly &
Some model parameters not yet ruled out; at least two LSST observables can be calculated consistently &
Quasi-linear calculations possible &
Models an interesting (but possibly speculative) physical possibility &
Useful constraints from at least one LSST observable \tabularnewline

\hline

4 &
May be unnatural, but have appealing features &
Some model parameters not yet ruled out; LSST observables cannot be calculated consistently &
Linear calculations possible &
Models features of a broad class of theories, but may not be physically interesting &
Mildly useful constraints (possibly when combined with other surveys) \tabularnewline

\hline
5&
Unnatural models (but some authors choose to work on them still) &
Model is already (practically) ruled out &
Only background calculations possible &
Little applicability beyond the specific theory &
Little chance of useful observational constraints  \tabularnewline 
\hline
\hline
\end{tabular}
\caption{\label{Table-MG-criteria}A table of grading and criteria used to rank and select modified gravity and dark energy models. A lower score indicates a more favourable model.}

\end{table}

\subsection{Models that were not selected}
\label{sec:notselected}

In this section, we provide a partial listing of the models that were not selected, along with brief justifications for these choices.

\begin{itemize}
\item{\it Quintessence (minimally-coupled)}:
This class of dark energy models includes the subclasses of freezing, thawing, tracker and Big Rip models, which correspond to specific classes of potential for the scalar field \citep{2005Caldwell-Linder-quintessence}. Theoretical tools are well-developed. Effects on perturbative observables are proportional to $1+\omega$, so are potentially small. This model is fully specified once $w(z)$ is known, {\color{black} so there is little to learn beyond the  core LSST analysis of (say) $\{w_0, w_a\}$ or other forms of $w(z)$CDM.} Minimally-coupled quintessence is a sub-class of Horndeski theories, so is automatically included in our priority model.

\item{\it k-essence}:
Similar to quintessence, but in this model the scalar field is permitted to have a non-standard kinetic term \citep{2000-Armendariz-Mukhanov-Steinhardt,2001-Armendariz-Mukhanov-Steinhardt}. Effects on perturbative observables are also proportional to $1+\omega$, so are potentially small.  It is also a sub-class of Horndeski, so is already included in our priority model.



\item{\it Brans-Dicke}: 
A classic scalar-tensor theory. It lacks a screening mechanism, so it is subject to Solar System constraints. Current constraints on the single free parameter of the model ($\omega$) are $\omega>40,000$ \citep{Cassini2003}, where $w\rightarrow \infty$ is the GR limit. This is also a sub-class of Horndeski, so is already included in our priority model.


\item{\it Ho\u{r}ava-Lifshitz}:
Orignally studied as a toy model of quantum gravity. The theory borrows ideas from condensed matter theory to introduce anisotropic scalings between space and time at high energies, that render the theory UV-safe (in contrast to the divergent behaviour of GR at high energies). See the original papers by \cite{HoravaTheory,Horava2010imp1} and also  the reviews by  \cite{Sotiriou2011HL,WangA2017HL}, and references therein. \cite{2016Frusciante-Horava-EFT} used the EFT of Dark Energy framework to find that Ho\u{r}ava-Lifschitz gravity strongly affects observables such as the matter power spectrum, weak lensing power spectra, ISW effect and CMB B-mode polarization. In turn, these translate into strong constraints on the theory's parameters. We consider these constraints to be sufficiently strong that Ho\u{r}ava-Lifschitz should not be a focus for our work. 

\item{\it DGP}:
Theoretical tools for DGP theories are well-developed, including N-body simulations, and theory has interesting braneworld-inspired motivations. However, the self-accelerating branch has a ghost, and hence is non-viable, see e.g.,  \citep{Luty2003DGP,Nicolis2004DGP,Gorbunov2006DGP,Charmousis2006DGP}. In addition, growth constraints have essentially ruled out this branch \citep{Fang2008DGP,Lombriser2009DGP}. Viable theories on the normal (non self-accelerating) DGP branch are also disfavored by growth data \citep{Barreira2016DGP,Hernandez2018DGP}.

\item{\it Einstein Gauss-Bonnet}:
This theory may have interesting connections to heterotic string theories \citep{2012-Clifton-MG}. In its original form, it is necessarily equivalent to GR in four dimensions, so not obviously testable by cosmological probes. Modified versions (e.g., f(G) models) have been explored, but are relatively constrained \citep{2007-Li-Barrow-Mota,2010-deFelice-Mota-Tsujikawa}. The theory can be mapped onto an instance of Horndeski gravity \citep{BertiElAlReview2015}, which is already included in our priority model.

\item{\it Kaluza-Klein}:
Originally posited as an attempt to unify gravity and electromagnetism, by considering a 4+1 manifold where one spatial dimension is compactified (see \cite{2012-Clifton-MG} and references therein). This leads to an effective 4D theory with an extra scalar field. Similar ideas later manifested in the compactifications of Calabi-Yau manifolds in string theory and string gas cosmology. {\color{black} A successful explanation of why some dimensions grow large, whilst others stay small, remains a major conceptual challenge for Kaluza-Klein theory. Current collider constraints bound the approximate size of the extra dimension to be $L\leq 10^{-19}$m; one might naturally expect it to be approximately Planckian in size.} Theoretical tools for Kaluza-Klein cosmology are not well-developed.

\item{\it MOND}:
Non-linear theoretical tools for MOND (modified Newtonian dynamics) are well-developed, but predictions for background and linear perturbations are unclear as no covariant Lagrangian formulation exists. It can be used as a toy model for studies around the galaxy scale, but not as a full cosmological theory of gravity. Dynamical predictions work well for galaxies \citep{Milgrom1983MOND,McGaugh2016MOND}, but are not successful for clusters \citep{Sanders1999MOND,Hodson2017MOND}.

\item{\it TeVeS}: Tensor-Vector-Scalar theory (TeVes) is a relativistic, covariant generalisation of MOND (see above), due to \cite{2004Bekenstein1,2004Benstein2}. It involves one each of additional tensor, vector and scalar fields, leading to a significantly complicated theory. Nevertheless, the cosmological linear perturbation theory of the model has been developed \citep{2009Skordis-TeVeS}, and can give rise to consistent LSS observations. However, the standard `vanilla' TeVeS model violates the bounds on the speed of gravitational waves from the GW170817 event. Extended TeVeS models consistent with the GW constraints exist, but the cosmological predictions of these are not well-developed, see for example \citep{Skordis2019}.

\item{\it Emergent spacetime}:
A term that has been to applied to several ideas \citep{2011Verlinde-emergent,2010Padmanabhan-emergent,1995Jacobson-emergent}. These have interesting physical motivations, attempting to explain gravity as a phenomenon emerging from a coarse-grained or thermodynamic `average' over a microscopic theory of spacetime. However, concrete and testable predictions are generally not yet well-developed. It is unclear that some of the most interesting aspects (e.g.,  entanglement structure of spacetime) will be testable.

\item{\it Einstein-Cartan-Sciama-Kibble} (ECSK) theory:
This theory is equivalent to GR, but with the assumption of zero torsion lifted. In ECSK theory, matter fields with spin act as a source of torsion for the metric, meaning that the metric connection is no longer symmetric (see \cite{RevModPhys1976-Einstein-Cartan,2006-Trautman-EC-theory}). Deviations from GR only become significant in very high spin-densities of matter, such as those potentially present in the early universe. Theories with torsion have been studied in a cosmological context in the hope that they could avoid early-time singularities; however, this only seems to work under very unrealistic early universe conditions.

\item{\it Conformal gravity}:
Conformal gravity replaces the Einstein-Hilbert action with the square of the Weyl tensor, leading to a higher-derivative theory \cite{Mannheim1,Mannheim3}. Work has been done to study the model as a theory of quantum gravity, and modifications to the gravitational potential have been claimed to eliminate the need for dark matter. However, there are serious debates about the stability of the theory and the presence of ghosts \cite{Mannheim4NG,Mannheim2NG,Pavsic2013,Pavsic2016}. Cosmological predictions are not well-developed.

\end{itemize}

\noindent {\bf Acknowledgments}
\\
%

M.I. acknowledges that this material is based upon work supported in part by NSF under grant AST-1517768 and the U.S. Department of Energy, Office of Science, under Award Number DE-SC0019206. T.B. acknowledges support from All Souls College, Oxford, and from the UK Royal Society. PGF is supported by the ERC and the Beecroft Trust. EL was supported in part by by the U.S. Department of Energy, Office of Science, Office of High Energy Physics, under Award DE-SC-0007867.

The DESC acknowledges ongoing support from the Institut National de Physique Nucl\'eaire et de Physique des Particules in France; the Science \& Technology Facilities Council in the United Kingdom; and the Department of Energy, the National Science Foundation, and the LSST Corporation in the United States.  DESC uses resources of the IN2P3 Computing Center (CC-IN2P3--Lyon/Villeurbanne - France) funded by the Centre National de la Recherche Scientifique; the National Energy Research Scientific Computing Center, a DOE Office of Science User Facility supported by the Office of Science of the U.S.\ Department of Energy under Contract No.\ DE-AC02-05CH11231; STFC DiRAC HPC Facilities, funded by UK BIS National E-infrastructure capital grants; and the UK particle physics grid, supported by the GridPP Collaboration.  This work was performed in part under DOE Contract DE-AC02-76SF00515.

\newpage

\appendix

\section{Linear evolution equations for Horndeski theories}
\label{app:horndeski}

In this appendix, we show the full system of linear evolution equations for the Horndeski class of scalar field models. These are taken from \cite{Bellini14} and use the restriction $\alpha_H=0$, as justified in the main text.

\subsection{linearized Einstein equations for Horndeski gravity}
We use the notation $G_{4X} = dG_4/dX$. The linearized field equations are (where on the LHS we indicate which tensorial component of the Einstein field equations these correspond to):
\begin{align}
00:\quad & 3\left(2-\alpha_{\textrm{B}}\right)H\dot{\Phi}+\left(6-\alpha_{\textrm{K}}-6\alpha_{\textrm{B}}\right)H^{2}\Psi+\frac{2k^{2}\Phi}{a^{2}}\label{eq:Hamilt}\\
 & -\left(\alpha_{\textrm{K}}+3\alpha_{\textrm{B}}\right)H^{2}\dot{v}_{X}-\left[\alpha_{\textrm{B}}\frac{k^{2}}{a^{2}}-3\dot{H}\alpha_{\textrm{B}}+3\left(2\dot{H}+\tilde{\rho}_{\text{m}}+\tilde{p}_{\text{m}}\right)\right]Hv_{X}=-\tilde{\rho}_{\text{m}}\delta_{\text{m}}\,,\nonumber \\
0i:\quad &2\dot{\Phi}+\left(2-\alpha_{\textrm{B}}\right)H\Psi-\alpha_{\textrm{B}}H\dot{v}_{X}-\left(2\dot{H}+\tilde{\rho}_{\text{m}}+\tilde{p}_{\text{m}}\right)v_{X}=-\left(\tilde{\rho}_{\text{m}}+\tilde{p}_{\text{m}}\right)v_{\text{m}}\label{eq:Momentum}\\
ij \;\mathrm{traceless}:\quad & \Psi-\left(1+\alpha_{\textrm{T}}\right)\Phi-\left(\alpha_{\textrm{M}}-\alpha_{\textrm{T}}\right)Hv_{X}=\tilde{p}_{\text{m}}\pi_{\text{m}}\label{eq:Aniso} \\
 \mathrm{trace}: \quad & 2\ddot{\Phi} -\alpha_{\textrm{B}}H\ddot{v}_{X}+2\left(3+\alpha_{\textrm{M}}\right)H\dot{\Phi}+\left(2-\alpha_{\textrm{B}}\right)H\dot{\Psi}\label{eq:PresEq}\\
 & +\left[H^{2}\left(2-\alpha_{\textrm{B}}\right)\left(3+\alpha_{\textrm{M}}\right)-\left(\alpha_{\textrm{B}}H\right)^{.}+2\dot{H}-\left(\tilde{\rho}_{\text{m}}+\tilde{p}_{\text{m}}\right)\right]\Psi\nonumber \\
 & -\left[\left(2\dot{H}+\tilde{\rho}_{\text{m}}+\tilde{p}_{\text{m}}\right)+\left(\alpha_{\textrm{B}}H\right)^{.}+H^{2}\alpha_{\textrm{B}}\left(3+\alpha_{\textrm{M}}\right)\right]\dot{v}_{X}\nonumber \\
 & -\left[2\ddot{H}+2\dot{H}H\left(3+\alpha_{\textrm{M}}\right)+\dot{\tilde{p}}_{\text{m}}+\alpha_{\text{M}}H\tilde{p}_{\text{m}}\right]v_{X}=\delta \tilde{p}_{\text{m}} \nonumber 
\end{align}
where $\tilde{p}_{\textrm{m}} = p_\textrm{m}/M_{\star}^2$ and $\tilde{\rho}_{\textrm{m}} = \rho_\textrm{m}/M_{\star}^2$. 

The evolution equation for the scalar field fluctuation $v_X$ reads
\begin{align}
3H \alpha_{\textrm{B}}\ddot{\Phi}&+H^{2}\alpha_{\textrm{K}}\ddot{v}_{X}-3\left[\left(2\dot{H}+\tilde{\rho}_{\text{m}}+\tilde{p}_{\text{m}}\right)-H^{2}\alpha_{\textrm{B}}\left(3+\alpha_{\textrm{M}}\right)-\left(\alpha_{\textrm{B}}H\right)^{.}\right]\dot{\Phi}\label{eq:scalar}\\
 & +\left(\alpha_{\textrm{K}}+3\alpha_{\textrm{B}}\right)H^{2}\dot{\Psi}-2\left(\alpha_{\textrm{M}}-\alpha_{\textrm{T}}\right)H\frac{k^{2}}{a^{2}}\Phi-\alpha_{\textrm{B}}H\frac{k^{2}}{a^{2}}\Psi-\nonumber \\
 & -\left[3\left(2\dot{H}+\tilde{\rho}_{\text{m}}+\tilde{p}_{\text{m}}\right)-\dot{H}(2\alpha_{\textrm{K}}+9\alpha_{\textrm{B}})-H\left(\dot{\alpha}_{\textrm{K}}+3\dot{\alpha}_{\textrm{B}}\right)-H^{2}\left(3+\alpha_{\textrm{M}}\right)\left(\alpha_{\textrm{K}}+3\alpha_{\textrm{B}}\right)\right]H\Psi+\nonumber \\
 & +\left[2\dot{H}\alpha_{\textrm{K}}+\dot{\alpha}_{\textrm{K}}H+H^{2}\alpha_{\textrm{K}}\left(3+\alpha_{\textrm{M}}\right)\right]H\dot{v}_{X}+H^{2}M^{2}v_{X}+\nonumber \\
 & +\left[-\left(2\dot{H}+\tilde{\rho}_{\text{m}}+\tilde{p}_{\text{m}}\right)+2H^{2}\left(\alpha_{\textrm{M}}-\alpha_{\textrm{T}}\right)+H^{2}\alpha_{\textrm{B}}\left(1+\alpha_{\textrm{M}}\right)+\left(\alpha_{\textrm{B}}H\right)^{.}\right]\frac{k^{2}}{a^{2}}v_{X}=0\,,\nonumber 
\end{align}
with
\begin{align}
H^{2}M^{2} & \equiv 3\dot{H}\left[\dot{H}\left(2-\alpha_{\textrm{B}}\right)+\tilde{\rho}_{\text{m}}+\tilde{p}_{\text{m}}-H\dot{\alpha}_{\textrm{B}}\right]-3H\alpha_{\textrm{B}}\left[\ddot{H}+\dot{H}H\left(3+\alpha_{\textrm{M}}\right)\right]\,.\label{eq:Mass}
\end{align}

\subsection{Growth equations for Horndeski gravity with pressureless matter}
\label{growth}

The full set of field equations above is somewhat messy. A more convenient system to solve can be formed by eliminating the scalar field perturbation from Eqns.~\ref{eq:Hamilt}, \ref{eq:Momentum} \& \ref{eq:Aniso} (note that this involves taking a time derivative of eq.\ref{eq:Aniso}). The result is a second order differential equation for the potential $\Phi$, sourced by $\delta_m$. This is, in fact, the first step in deriving the quasi-static approximation expressions, as described in \S\ref{sec:horndeski:implementation}.

We will also make the simplification of a pressureless matter sector, setting $p_m=\pi_m=0$ (where $\pi_m$ here is the anisotropic stress of matter). Furthermore, Bellini et al. \cite{Bellini14} argue that the velocity perturbation $v_m$ can be neglected on subhorizon scales. Implementing these simplifications:

\begin{align}
&\ddot \Phi +  \frac{\beta_1 \beta_2 + \beta_3  \alpha_B^2 \,\tilde k^2}{\beta_1+\alpha_B^2 \tilde k^2} H \dot \Phi+   \frac{ \beta_1 \beta_4+\beta_1 \beta_5  \,\tilde k^2   +c_s^2  \alpha_B^2 \tilde k^4} {\beta_1+ \alpha_B^2 \,\tilde k^2 }  H^2  \Phi = - \frac{1}{2M_*^2} \Bigg[  \frac{\beta_1\beta_6 + \beta_7  \alpha_B^2 \, \tilde k^2  }{\beta_1+ \alpha_B^2 \,\tilde k^2  }  \delta \rho_{\rm m} \Bigg]\;, \label{Phi_evol_cdm}
\end{align}
and the slip relation:
\begin{align}
 &\alpha_{\textrm{B}}^{2}\frac{k^{2}}{a^{2}}  \left[\Psi-\Phi\left(1+\alpha_{\textrm{T}}+\frac{2\left(\alpha_{\textrm{M}}-\alpha_{\textrm{T}}\right)}{\alpha_{\textrm{B}}}\right)\right]+ \beta_{1}\left[\Psi-\Phi\left(1+\alpha_{\textrm{T}}\right)\left(1-\frac{2\alpha H^{2}\left(\alpha_{\textrm{M}}-\alpha_{\textrm{T}}\right)}{\beta_{1}}\right)\right]\nonumber \\
 &\quad\quad\quad=\left(\alpha_{\textrm{M}}-\alpha_{\textrm{T}}\right)\left[\alpha_{\textrm{B}}\frac{{\rho}_{\textrm{m}}\delta_{\textrm{m}}}{M_*^2}-2H\alpha\,\dot{\Phi}\right]\,. \label{eq:anisotropy}
\end{align}
where ${\tilde{k}=k/aH}$ and the $\beta_i$ functions are given below. Bellini et al. comment that whenever $\alpha_B\neq 0$, the slip relation contains a `braiding scale' at which the behaviour of the perturbations change in their dynamical behaviour. However, it is possible that the braiding scale is at or above the horizon, in which case we would not be sensitive to the transition.

The system is closed by the standard evolution equations for CDM:
\begin{align}
\dot{\delta}_{\text{m}}-\frac{k^{2}}{a^{2}}v_{\text{m}}=3\dot{\Phi}\,,\qquad\dot{v}_{\text{m}}=-\Psi\,,\label{eq:MatterCons}
\end{align}
Eqns.~\ref{Phi_evol_cdm} - \ref{eq:MatterCons} can then be coded up to solve for the evolution of $\delta_m$, from which the growth rate can be computed.
 
The $\beta_i$ functions are:
\begin{align}
\beta_1  &  = - \alpha_K \frac{\tilde{\rho_{\rm m}}}{ H^2} - 2 \DD \left( \frac{\dot H}{H^2} + \alpha_T -  \alpha_M   \right)  \;, \label{beta1}\\
\beta_2 & \equiv 2(2+\alpha_M) + 3 \Upsilon  \;, \\
\beta_3 & \equiv 3+\alpha_M + \frac{\alpha_B^2}{H  \DD} \left(  \frac{ \alpha_K}{\alpha_B^2} \right)^{\hbox{$\cdot$}} \;,\\
\beta_4 & \equiv (1+\alpha_T) \big[ 2 \dot H /H^2 + 3 (1+ \Upsilon) + \alpha_M \big] + \dot \alpha_T /H \;,  \\
\beta_5 & \equiv c_s^2 - \frac{ 2 \alpha_B (\beta_3- \beta_2)}{\DD}
+\frac{\alpha_B^2 }{4\beta_1} (1+\alpha_T) (\beta_3 - \beta_2 ) +\frac{\alpha_B^2 \beta_4}{4\beta_1}
 \;, \\
\beta_6 & \equiv \beta_7 + \frac{\alpha_B(\beta_3 -\beta_2)}{\DD} \;, \\
\beta_7 & \equiv c_s^2 +  \frac{\alpha_B^2/2 (1+ \alpha_T ) -  \alpha_B (\alpha_T - \alpha_M)}{ \DD} \;,  \\
\beta_8 & \equiv \beta_9 -\frac{(\alpha_K +3 \alpha_B )(\beta_3 -\beta_2)}{\DD} \;, \\
\beta_9 & \equiv   -(1+3 c_s^2 +  \alpha_T)  + \frac{\alpha_B^2}{H  \DD} \left(  \frac{ \alpha_K}{\alpha_B^2} \right)^{\hbox{$\cdot$}} \;, \\
\beta_{10} & \equiv  - 6 (1+ \Upsilon) - 4 \dot H/H^2 \;, \\
\beta_{11} & \equiv\frac{2}{3} -  \frac{\alpha_B^2}{2\beta_1}\big[ (2 - \alpha_M) +2  \dot H/H^2  \big] -  \frac{\alpha_B^4}{ 2\beta_1 H \DD} \left(  \frac{ \alpha_K}{\alpha_B^2} \right)^{\hbox{$\cdot$}} \; \\
\gamma_1 &\equiv \alpha_K \frac{\rho_D+p_D}{4 H^2 M_*^2 } -3 \alpha_B^2 \frac{\dot H}{H^2}  \;, \label{gamma1} \\
\gamma_9 & \equiv \DD\frac{\alpha_T-\alpha_M}{2 }\; . 
\end{align}
with
\begin{align}
\alpha&=\alpha_K+\frac{3}{2}\alpha_B^2 \label{alpha-again}\\
12 \beta_1 H^3 M_*^2 \Upsilon &\equiv   \ 2 {\DD} M_*^2  \Big\{ \big[ \dot H +  (\alpha_T - \alpha_M) H^2 \big]^{\hbox{$\cdot$}} +  (3+\alpha_M) H \big[ \dot H +  (\alpha_T - \alpha_M) H^2 \big]  \Big\} \\
& - \rho_{\rm m}  H (\alpha_K +3 \alpha_B) (\alpha_T - \alpha_M) + \frac{3}{2}\rho_{\rm m}  \frac{\alpha_B^4}{  \DD} \left(  \frac{ \alpha_K}{\alpha_B^2} \right)^{\hbox{$\cdot$}}    \\
c_s^2 &= - \frac{(2-\alpha_B) \Big[ \dot H -  (\alpha_M - \alpha_T) H^2 - H^2 \alpha_B/2 (1+\alpha_T)\Big] - H \dot \alpha_B + \tilde{\rho}_{\rm m}}{H^2 \DD}\label{cs2-again}
\end{align}

\section{Linear perturbation equations for the $RT$ non-local model}

Using again a prime to denote the derivative with respect to $x=\ln a$  defining the variables $V=H_0a^{-1}S_0$ and $Z=H_0^2 S$, the perturbed field equations are given by \cite{Dirian2014CP,Nesseris2014NL} 
\bees
&&\hspace{-8mm}\hat{k}^2 \Phi + 3(  \Phi' - \Psi) = \frac{3}{2h^2\rho_0}
\[ \delta \rho + \gamma \rho_0 \( \delta U - h  \delta V' + 2 h \Psi \bar{V}' + h  \Psi' \bar{V} \) \], \label{EE1xRT}\\
&&\hspace{-8mm} \hat{k}^2 (\Phi' - \Psi ) = - \frac{3}{2h^2\rho_0}
\[ \bar{\rho} (1+w) \hat{\theta} + \hat{k}^2 \gamma \rho_0 
\(  h^2 \delta Z - \frac{h^2}{2} \delta Z' + h \Psi \bar{V} - \frac{h}{2} \delta V \) \], \label{EE2xRT} \\
&&\hspace{-8mm} \hat{k}^2 (\Psi + \Phi) =  \frac{9 }{2h^2\rho_0} 
\[ \bar{\rho} (1+w) e^{2x}\sigma +  \frac{2}{3} \hat{k}^2 \gamma\rho_0  h^2 \delta Z  \],  \label{EE3xRT}\\
&& \hspace{-8mm} \Phi'' + (3+\zeta)  \Phi' -  \Psi' - (3+2\zeta) \Psi + \frac{\hat{k}^2}{3} (\Phi + \Psi) \nonumber \\
&& = - \frac{3}{2h^2\rho_0}
\[ \delta p - \gamma \rho_0 \( \delta U - h (  \Phi' - 2 \Psi ) \bar{V} - h \delta V - \frac{\hat{k}^2}{3} h^2 \delta Z \)  \]\, . \label{EE4xRT}
\ees
Next, the linearized equations for the auxiliary fields read
\bees
&& \hspace{-10mm} \delta U'' + (3 + \zeta)  \delta U' + \hat{k}^2 \delta U = 2 \hat{k}^2 (\Psi + 2 \Phi)   + 6 ( \Phi'' +(4+\zeta)  \Phi' ) - 6 \[ \Psi' + 2(2+ \zeta) \Psi \] \nonumber \\
&& \hspace{5mm}
+ 2 \Psi  \bar{U}'' +
\[ 2 \Psi (3+\zeta) + (\Psi' - 3  \Phi') \]  \bar{U}'\,\label{AppU} \\
&&\hspace{-10mm} \delta V'' + (3+\zeta)  \delta V' 
+ \frac{\hat{k}^2}{2} h (  \delta Z' - 4 \delta Z )- h^{-1} \d U'  =  2 \Psi  \bar{V}''
 + \[ 2(3+\zeta) \Psi + 3( \Psi' -  \Phi') \]  \bar{V}' \nonumber \\
&&  \hspace{5mm}
 + \[ \Psi'' + (3 + \zeta)  \Psi' + 6 \Phi' \] \bar{V}  - \[ (1/2) \hat{k}^2 - 3 \] \( \delta V - 2 \Psi \bar{V} \)\, ,
\label{AppV} \\
&& \hspace{-10mm}
 \delta Z'' + (1+\zeta)\delta Z' + 2 \(  \hat{k}^2 - (3+\zeta) \) \delta Z =2h^{-2}\d U\nn\\
&&\hspace{5mm}-h^{-1}\[  \delta V' + 5 \delta V - 4 \Psi  \bar{V}' - 2 (  \Psi' -  \Phi' + 4 \Psi ) \bar{V} \]
\label{AppZ}\, .
\ees

Finally, the perturbed Einstein equations can again be recast in the form of an effective fluid as 
\begin{align}
\delta \rho_{\rm DE}& \equiv \gamma \rho_0 \big( \delta U - h  \delta V' + 2 h \Psi  \bar{V}' + h\bar{V}  \Psi'  \big), \\
\bar{\rho}_{\rm DE} (1+w_{\rm DE}) \hat{\theta}_{\rm DE} &\equiv \hat{k}^2 \gamma \rho_0 \bigg(  h^2 \delta Z - \frac{h^2}{2} \delta Z' + h \Psi \bar{V} - \frac{h}{2} \delta V \bigg),\\
\bar{\rho}_{\rm DE} (1+w_{\rm DE}) \sigma_{\rm DE} &\equiv  \frac{2}{3} \hat{k}^2 \gamma\rho_0 e^{-2x} h^2 \delta Z, \\
\delta p_{\rm DE} &\equiv  - \gamma \rho_0 \bigg( \delta U - h (  \Phi' - 2 \Psi ) \bar{V} - h \delta V - \frac{\hat{k}^2}{3} h^2 \delta Z \bigg)\, .
\end{align}

\newpage
\bibliography{LSST_DESC_BwCDM}

\end{document}